\documentclass[transmag]{IEEEtran}
\usepackage{latexsym}
\usepackage{graphicx}
\usepackage{amsfonts,amssymb,amsmath}
\usepackage{hyperref}
\usepackage{multirow,makecell}

\newcommand{\norm}[1]{\|#1\| }

\setcellgapes{1pt}
\makegapedcells
\newcolumntype{C}[1]{>{\centering \arraybackslash }b{#1}}

\def\BibTeX{{\rm B\kern-.05em{\sc i\kern-.025em b}\kern-.08em T\kern-.1667em\lower.7ex\hbox{E}\kern-.125emX}}
\begin{document}

\title{Polarization-Based Reconfigurable Tags for Robust Ambient Backscatter Communications}

\author{R. Fara, 
D.-T. Phan-Huy, 
\IEEEmembership{Member, IEEE}, 
A. Ourir, 
Y. Kokar, 
J.-C. Prévotet, 
M. Hélard,
\IEEEmembership{Member, IEEE},
\\
M. Di Renzo,
\IEEEmembership{Fellow, IEEE}
and J. de Rosny,
\thanks{This work has been partially supported by the French Project ANR Spatial Modulation under grant ANR-15-CE25-0016 (https://spatialmodulation.eurestools.eu/). This paper is an extended version of the conference paper \cite{ref12}.}
\thanks{R. Fara is with Universit\'e Paris-Saclay, CNRS and CentraleSup\'elec, Laboratoire des Signaux et Syst\`emes,  91192 Gif-sur-Yvette, and Orange Labs Networks, Châtillon, France (e-mail: romain.fara@orange.com).} 
\thanks{D.-T. Phan-Huy is with Orange Labs Networks, Châtillon, France (e-mail: dinh-thuy.phan-huy@orange.com).}
\thanks{A. Ourir is with ESPCI Paris, PSL University, CNRS, Institut Langevin, Paris, France.}
\thanks{Y. Kokar is with Univ. Rennes, INSA Rennes, IETR, CNRS, UMR 6164, Rennes, France}
\thanks{J.-C. Prévotet is with Univ. Rennes, INSA Rennes, IETR, CNRS, UMR 6164, Rennes, France}
\thanks{M. Hélard is with Univ. Rennes, INSA Rennes, IETR, CNRS, UMR 6164, Rennes, France}
\thanks{M. Di Renzo is with Universit\'e Paris-Saclay, CNRS and CentraleSup\'elec, Laboratoire des Signaux et Syst\`emes,  91192 Gif-sur-Yvette, France (email: marco.direnzo@centralesupelec.fr).}
\thanks{J. de Rosny is with ESPCI Paris, PSL University, CNRS, Institut Langevin, Paris, France.
}
}

\IEEEtitleabstractindextext{\begin{abstract}Ambient backscatter communication is an emerging and promising low-energy technology for the Internet of Things. In such a system, a tag sends a binary message to a reader by backscattering a radio frequency signal generated by an ambient source. The tag can operate without battery and without generating additional radio waves. However, the tag-to-reader link suffers from the source-to-reader interference. In this paper, we propose a polarization-based reconfigurable antenna in order to improve the robustness of the tag-to-reader link against the source-to-reader direct interference. More precisely, we compare different types of tags’ antennas, different tags’ encoding schemes, and different detectors at the reader. By using analysis, numerical simulations, and experiments, we show that a polarization-based reconfigurable tag with four polarization directions significantly outperforms a non-reconfigurable tag, and provides almost the same performance as an ideal reconfigurable tag with a large number of reconfigurable polarization patterns.\end{abstract}

}

\maketitle

\section{Introduction}
\label{sectionI}
\IEEEPARstart{T}{he} recent development of the Internet of Things (IoT) has massively increased the number of connected devices. At the same time, despite the energy efficiency improvement brought by every mobile network generation, the energy consumption of wireless communications keeps increasing due to the fast growth of the number of devices \cite{ref1}.

Recently, the ambient backscatter (AmB) principle \cite{ref2} has been proposed for low-energy consumption communication. In an AmB system, a Radio Frequency (RF) tag transmits a binary message to an RF reader without a battery and without generating additional waves. The tag can be illuminated by an RF ambient source (such as a TV tower, a WiFi hot-spot or a 5G base station). In simple terms, the tag switches between two states: a backscattering state during which it backscatters the ambient signal, and a transparent state during which it has a weaker effect on the ambient signal. The two distinct states constitute a code for the bits “1” and “0”, respectively. The simplest implementation for a tag is constituted by a dipole antenna that switches between two different load impedances: a null impedance (the two branches of the dipole are short-circuited) or an infinite impedance (the two branches of the dipole are open-circuited). Such tag does not generate any additional RF waves and can therefore operate without battery. An energy harvesting device is then sufficient to power an RF switch and a low power micro-controller. On the RF reader side, the simplest receiver is an energy detector that compares the received power level with a threshold (the time-windowed received power for instance) to determine the state of the tag and to detect the transmitted bits. The performance of the energy-detector increases with the signal contrast, i.e., with the difference between the signals’ amplitude during the first state and the second state.

Due to the low energy consumption, AmB has been identified as a promising technology for application to the IoT \cite{ref3}. However, the tag-to-reader link suffers from the source-to-reader direct interference in many different ways. Firstly, on average, the tag-to-reader signal is weak compared to the source-to-reader direct signal. This is due to the fact that a fraction of the incident ambient signal is backscattered and diffused in many directions. Secondly, a tag in a deep fade of the ambient signal is invisible to the reader. This may happen, typically, in a rich scattering environment. Thirdly, even in line-of-sight (LOS) and even if the tag-to-reader signal is strong, the signal-to-noise-ratio (SNR) contrast can be close to zero for some locations of the tag. Indeed, there are locations where the tag-to-reader signal and the source-to-reader direct interference interfere together so that the received signals in the two distinct states are different in phase but are equal in amplitude. In these locations, the signal contrast is null and the energy detector performance is poor. The authors of \cite{ref4} show, by using simulation and experiments, that the signal contrast experiences deep fades when the tag is located on ellipses that have the source and the reader as foci. These ellipses are regularly spaced by half a wavelength. Such deep fades of the signal contrast may also occur in scattering environments. Even if the tag is not in such locations, in addition, the performance may not be good enough if the polarizations of the tag, reader, and source are not matched \cite{ref5}. 

Channel polarization has been studied extensively for radio frequency identification (RFID) tags and RFID readers in \cite{ref6}. To guarantee a minimum polarization match, a circular polarized antenna at the reader side has been proposed in \cite{ref7}. However, ambient sources may not use such type of antennas systematically.

More recently, a tag based on a compact reconfigurable antenna has been proposed to achieve higher data rates for application to AmB systems \cite{ref8}. The antenna switches between four radiation patterns with distinct dominant linear polarizations. Such compact polarization-based reconfigurable (PR) antennas have been proposed for spatial modulation application \cite{ref9}, \cite{ref10}. These PR antennas are used in \cite{ref9} and \cite{ref10} to spatially modulate the transmitted signal thanks to different patterns with low inter-correlation. In our case, we propose to use these PR antennas to backscatter the ambient signal and to improve the robustness of AmB communications as well.
 
\begin{figure}
\centerline{\includegraphics[width=3.5in]{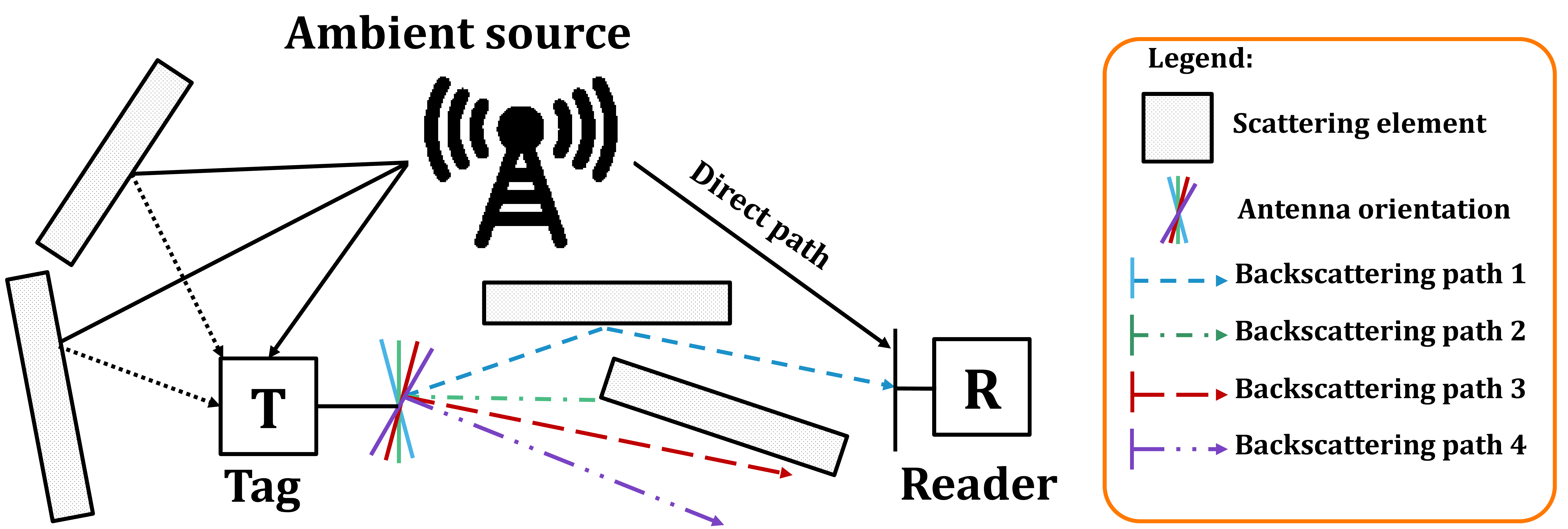}}
\caption{Tag with different polarizations that is illuminated by an ambient signal in a rich scattering environment, and that backscatters its message to the reader.\label{fig1}}
\end{figure}
This PR tag attains $log_{2}(4)=2$ bits per switching period instead of 1 bit per switching period (that would be obtained with a 2-state tag).

Other types of detectors have been implemented for some applications of ambient backscattering, such as solutions based on the minimum mean square error instead of the conventional energy detector \cite{ref11}. These solutions are, however, complex, and hence we propose to study the least square error estimator that has lower complexity.

In this paper, we introduce a PR tag to improve the robustness of the tag-to-reader link against the source-to-reader direct interference. As illustrated in Fig. \ref{fig1}, we consider a PR tag that is able to switch between several polarizations. Instead of using the PR tag to increase the spectral efficiency and the data rate of the tag-to-reader link \cite{ref8}, we use it to improve the robustness of the tag-to-reader link against noise and source-to-reader interference. 

In addition, we consider and compare two different coding schemes that exploit polarization diversity to improve the performance. In the first considered coding scheme, the tag, as any standard two-state tag, sends its message by switching its operation between a backscattering state and a transparent state. In practice, this is realized by switching the antenna between two different load impedances. This coding scheme is referred to as “\textit{Backscattering/Transparent Scheme}" (BTS). With this approach, however, the tag sends the same message several times, by using a different configured radiation pattern (and corresponding different polarization) at each transmission time. The rationale of this approach lies in the possibility of receiving a good quality signal over at least one polarization state. 

To make an initial assessment of the benefits of the proposed tag, we introduce a simple model for the tag that can be studied theoretically and numerically with the aid of an electromagnetics-based simulation software (4NEC2) based on the Method of Moments (MoM), and that can be tested experimentally by using dipole antennas. We model the tag as a device that can switch between $N_{pol}$ radiation and corresponding polarization patterns. This is obtained by mechanically rotating a dipole antenna across $N_{pol}$ different orientations. The dipole is connected to a load impedance with two different values in order to create the backscattering and the transparent transmission states. 

In addition, we introduce a second coding scheme, according to which the tag backscatters ambient signals with different polarization patterns. The information bits are encoded by switching among a subset of available patterns, according to the spatial modulation principle \cite{ref13}-\cite{ref19}. This coding scheme is referred to as “\textit{Polarization Coding Scheme}” (PCS). Similar to the BTS scheme, we expect that an appropriate (optimized) choice of the subset of polarization states will improve the system performance. The proposed implementation is assessed experimentally by using a compact reconfigurable antenna that is made of a split ring resonator antenna and a cross-polarization antenna. 

The rest of paper is organized as follow: Section \ref{sectionII} introduces the system model and the methodology for performance evaluation. Section \ref{sectionIII} reports a theoretical and simulation based performance analysis of the BTS coding scheme. In particular, the best direction of polarization is identified under LOS propagation conditions. Section \ref{sectionIV} reports the simulation and experimental validation of the proposed approach under a rich scattering propagation environment. Section \ref{sectionV} reports a performance comparison between an energy detector and a least square error receiver. Section \ref{sectionVI} reports simulation and experimental results that compare different compact reconfigurable antennas. Finally, Section \ref{sectionVII} concludes this paper.

\section{System Model and Methodology of Analysis}
\label{sectionII}
\subsection{System Model}
\label{sectionIIA}
We consider a system that consists of a source, a tag and a reader. The source radiates an ambient signal at the carrier frequency $f$ corresponding to the wavelength $\lambda$. The source and reader are equipped with a half-wavelength linearly polarized dipole antenna of length $l^D=\lambda/2$. We neglect the radius of the conductors (thin-wire approximation).

The positions and orientations of the antennas of the source, tag, and reader are denoted by ($x^S$,$y^S$,$z^S$), ($x^T$,$y^T$,$z^T$) and ($x^R$,$y^R$,$z^R$), respectively. They correspond to the Cartesian coordinates of the centers of the dipole antennas of the source, tag, and reader. The orientation angles of the dipole antennas of the source and reader are denoted by ($\phi^S$,$\phi^S$) and ($\phi^R$,$\phi^R$), respectively.

The antenna of the tag can switch among $N_{pol}$ radiation patterns with distinct dominant polarization states. 

The signal received by the reader is $y\in \mathbb{C}$:
\begin{equation}
y=(h^{SR}+\gamma h^{TR})\sqrt{P^{Tx}}+v,
\label{eq1}
\end{equation}
where $h^{SR}\in \mathbb{C}$ is the channel coefficient of the direct path between the source and the reader, $h^{TR}\in \mathbb{C}$ is the channel coefficient of the backscattered link between the source and the reader (i.e., the signal backscattered by the tag antenna),  $\gamma$ is a modulation factor that corresponds to the state of the tag (i.e., transparent mode or backscattering mode), $v\in \mathbb{C}$, is the Gaussian noise at the reader and $P^{Tx}$ is the power transmitted by the source. By introducing the aggregate channel $g=(h^{SR}+\gamma h^{TR})$, with $g\in \mathbb{C}$, the received signal can be rewritten as follows:
\begin{equation}
y=g\sqrt{P^{Tx}}+v.
\label{eq2}
\end{equation}

Let $P^{noise}=\mathbb{E}\big[\norm{v}^2\big]$ denote the received noise power, where $\norm{\cdot}^2$ is the Frobenius norm. The signal-to-noise ratio, denoted by $SNR^{Tx}$, is defined as the ratio between the power transmitted by the source and the noise power at the reader:
\begin{equation}
SNR^{Tx}=\frac{P^{Tx}}{P^{noise}}.
\label{eq3}
\end{equation}

To allow a fair comparison between different types of tags, the transmitted power and the noise power are assumed to be fixed. Hence, $SNR^{Tx}$ is assumed to be fixed. Also, we introduce $SNR^{captured}$ to denote the average signal-to-noise ratio (SNR), which is given by the ratio of the power captured by the reader ($P^{Rx}$) when the tag operates in the transparent mode and the noise power at the reader:
\begin{equation}
SNR^{captured}=\frac{P^{Rx}}{P^{noise}}.
\label{eq4}
\end{equation}

We consider a sequence of R bits $\mathbf{b}=b_1\dots b_r\dots b_R$ sent by the tag and denote by $\widehat{\mathbf{b}}=\widehat{b}_1\dots\widehat{b}_r \dots \widehat{b}_R$ the sequence of R bits detected by the reader. The Bit Error Rate (BER) is defined as follows
\begin{equation}
BER=\frac{\sum_{r=1}^R|\widehat{\mathbf{b}}_r-\mathbf{b}_r|^2}{R}.
\label{eq5}
\end{equation}

We consider that the target quality of service is met by the tag-to-reader link when $BER^{target}$ is achieved. The outage probability is hence defined, for given values of $SNR^{Tx}$ or $SNR^{captured}$,  as the probability of the event $BER<BER^{target}$.

\subsection{Propagation Environment}
\label{sectionIIB}
Different propagation environments are analyzed.
\begin{enumerate}
    \item \textit{LOS}. The LOS environment corresponds to a free space propagation environment between the source, the tag, and the reader.
    \item \textit{Scattering Environment}. In this case, we consider different elements including scatterers and reflectors. A scatterer is modeled as a conductive line defined by its length $l^{SC}$. The number of scatterers is denoted by $N^{SC}$. Reflectors are modeled as reflective planes, such as the ground plane. The number of reflectors is denoted by $N^{RP}$. Each scatterer is randomly positioned but the distance between any dipole antenna and the scatterer is $D^{SC-X}>\lambda$. The distance from each randomly distributed scatterer and the reader is constrained to be $D^{SC-R}<10\lambda$. Dipoles antennas are spaced at least half of the wavelength apart. As far as the experiments are concerned, a reverberation chamber is employed in order to reproduce realistic multipath propagation and a scattering environments.
\end{enumerate}

\subsection{Reconfigurable Antennas for the Tag}
\label{sectionIIC}
\begin{figure}
\centerline{\includegraphics[width=3.5in]{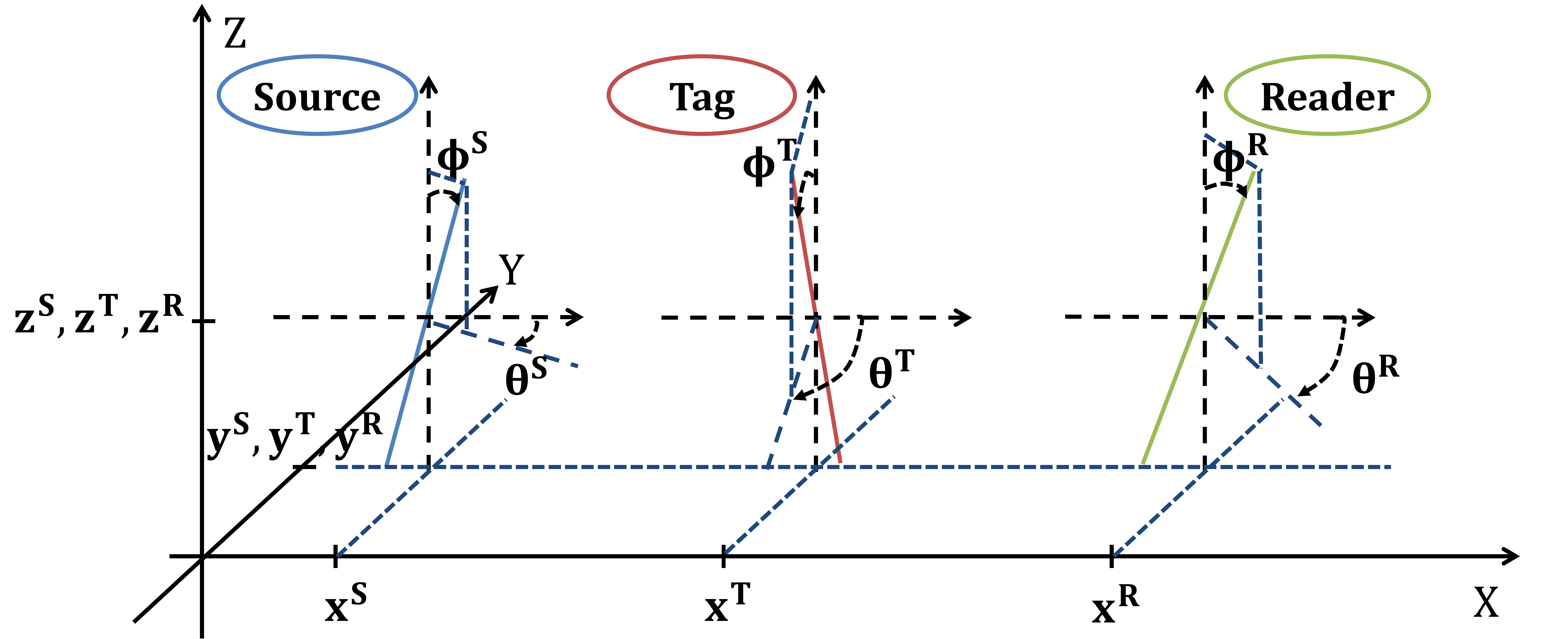}}
\caption{3D model of the system for a dipole tag.
\label{fig2}}
\end{figure}

In order to implement the tag, we consider three types of reconfigurable antennas.
\begin{enumerate}
    \item \textit{Rotating Dipole Antenna}. The tag is a mechanically rotating half-wavelength dipole antenna with linear polarization. By changing the orientation of the antenna, the polarization and the main orientation of the backscattering signal can be controlled and changed. This implementation allows us to have an antenna that is reconfigurable in polarization. More specifically, three types of tags are considered: 1) an “ideal” polarization reconfigurable (IPR) tag that is able to be configured for any directions of polarization; 2) a more realistic 4-polarization reconfigurable (4PR) tag that can realize four directions of polarization; and 3) a non-reconfigurable (NR) tag with a fixed direction of polarization. 
    
    As an illustrative example, Fig. \ref{fig2} shows the orientation angles of a tag. The angles are denoted by ($\phi^T$,$\phi^T$).

    \item \textit{Split Ring Resonators (SRR) Antenna}. In this case, the tag is realized by using split ring resonators \cite{ref8}, \cite{ref9}. More precisely, the tag is composed of four resonators and four PIN diodes. By controlling the PIN diodes, one can connect or disconnect the resonators and, hence, change the polarization pattern of the SRR antenna. In this paper, we consider only four patterns in order to compare the SRR antenna with the 4PR rotating dipole antenna. For performance optimization, the four patterns are those that provide the most distinguishable polarization states. The polarization patterns are provided in Section \ref{sectionVI} (Fig. \ref{fig19}).
    
    \item \textit{Cross-Polarization Antenna (XPOL)}. In this case, the tag is realized by using a cross-polarization antenna \cite{ref10}. The polarization state is changed by using PIN diodes. In particular, only four polarization states out of the eight possible polarization states are considered \cite{ref8}. The polarization patterns are provided in Section \ref{sectionVI} (Fig. \ref{fig20}).
\end{enumerate}

\subsection{Coding Schemes}
\label{sectionIID}
To transmit in a reliable manner the R bits from the tag to the reader, two coding schemes are analyzed.
\begin{enumerate}
    \item \textit{Backscattering/Transparent Scheme}. The tag sends its message, for a given polarization of the antenna, by switching between two states corresponding to different load impedances connected to the antenna. The two states are (i) the backscattering state that is obtained by short-circuiting the dipole antenna strands, so that the corresponding impedance is close to zero; and (ii) the transparent state that is obtained by open-circuiting the dipole antenna strands, so that the corresponding impedance is close to infinity. These two states have a different impact on the propagation of the signal.
    
    The BTS coding scheme corresponds to a well-used coding scheme in AmB communications. The backscattering state encodes bit 1 and the transparent state encodes bit 0. In \ref{eq1}, the backscattering state corresponds to a modulation factor $\gamma^{ON}=1$, i.e., $g^{(1)}=h^{SR}+h^{TR}$, and the transparent state corresponds to a modulation factor $\gamma^{OFF}=0$, i.e., $g^{(0)}=h^{SR}$. According to the BTS coding scheme, the tag repeats the message several times by using a different radiation pattern and polarization at each transmission instance in order to make the transmission more robust and reliable. It is expected, in fact, that at least one polarization pattern is more reliably detected by the reader.
    
    \item \textit{Polarization Coding Scheme}. The tag sends the message by switching between two polarization patterns of a pair of patterns (among several available pairs of patterns) and each pattern corresponds to a different bit. Similar to spatial modulation \cite{ref17}, the reader is able to determine the polarization pattern used by the tag. In this coding scheme, the tag is always backscattering but with different polarization patterns In particular, the bit 1 is encoded by one polarization pattern and the bit 0 is encoded by a different polarization pattern. In \ref{eq1}, the bit 1 corresponds to $g^{(1)}=h^{SR}+h^{TR(1)}$, and the bit 0 corresponds to $g^{(0)}=h^{SR}+h^{TR(0)}$. It is expected that this scheme provides good performance if the detector can easily distinct the two states of the tag, since it takes advantage of uncorrelated polarization patterns.
\end{enumerate}

\subsection{Detectors}
\label{sectionIIE}
At the reader, two different detectors are analyzed.
\begin{enumerate}
    \item \textit{Energy Detector (ED)}. An ED measures the voltage at the port of the dipole antenna, which is induced by the total received signal. The ED, in particular, determines whether the received average power is above or below a present threshold in order to detect the transmitted bits. The voltages measured at the dipole antenna port of the reader when the tag is backscattering and when the tag is transparent are denoted by $V^{ON}$ and $V^{OFF}$, respectively. From the measured voltages, the corresponding signal amplitudes $A^{ON}$ and $A^{OFF}$ can be deduced for each state. The signal contrast $\Delta A$ is defined as the difference of the received signals between the two states of the tag:
    \begin{equation}
        \Delta A=|A^{ON}-A^{OFF}|.
        \label{eq6}
    \end{equation}
    The receiver noise amplitude level is denoted by $A^{noise}$. Using this notation, the BER is defined as follows \cite{ref4}:
    \begin{equation}
        BER=0.5 \ erfc\Big(\frac{\Delta A}{A^{noise}}\Big).
        \label{eq7}
    \end{equation}
    where $erfc()$ denotes the complementary error function.
    
    \item \textit{Least Square Error Detector (LSE)}. This detector operates in two phases. First it estimates the channel for the states of the tag based on the coding scheme. Then, during the transmission of the tag, the transmitted bits are detected. The output of the detector can be formulated as follows:
    \begin{equation}
        \widehat{y}= arg \ min \ \Big\{ \norm{z-y}^2  \Big\} \ \text{with} \ z \in \big\{ g^{(0)},g^{(1)} \big\},
        \label{eq8}
    \end{equation}
   where $\widehat{y}$ is the estimate of $y$. For each bit $b_r$, the reader calculates $\widehat{y}$ for the two states $g^{(0)}$   and $g^{(1)}$ that correspond to the bit 0 and bit 1, respectively. The estimated bit $\widehat{b}_r$ corresponds to the smallest $\widehat{y}$.  

\end{enumerate}

\subsection{Channel Modeling}
\label{sectionIIF}
Three different channel models are considered, and are analyzed analytically, and by using simulations and experiments.
\begin{enumerate}
    \item \textit{Analytical Model}. In order to assess the performance of the considered backscattering system, we introduce a model for the polarization-based dipole antenna in a LOS propagation environment. This is discussed in Section \ref{sectionIII}.
    
    \item \textit{Numerical Electromagnetic Code (NEC) Simulator}. The dipole antennas are studied by using the 4NEC2 simulation software. 4NEC2 is a software based on the method of moments. By using 4NEC2, it is possible to model and configure the dipole antennas, the source, the tag, the reader, the $N^{SC}$ scatterers, the ground plane, the load impedances, and the physical characteristics of the dipole antennas and the conductive lines. In particular, this tool allows us to (i) configure the state of the tag (transparent or backscattering), by changing the load impedance; and (ii) to measure the resulting voltage at the reader port.
    
    \item \textit{Experiments}. Besides analysis and simulations, we perform measurements to estimate the channels. The obtained empirical channels are used to validate the analytical model and the simulations.
\end{enumerate}

\subsection{Summary of the Case Studies Analyzed}
\label{sectionIIG}
Table \ref{tab1} lists all systems and performance evaluation methodologies considered in this paper. We note that the XPOL and SRR antennas radiation patterns are complex to be modeled analytically and through 4NEC2. Therefore, these cases are analyzed only through experiments.

\begin{table}
\caption{Case Studied and Evaluation Methodologies}
\label{table}
\setlength{\tabcolsep}{3pt}
\begin{tabular}{|C{40pt}|C{44pt}|C{35pt}|C{28pt}|C{28pt}|C{40pt}|}
\hline
\multirow{2}{*}{\textit{\textbf{Section}}}&\multirow{2}{*}{\textit{\textbf{Environment}}}&\multirow{2}{*}{\textit{\textbf{Antenna}}}&\textit{\textbf{Coding}}&\multirow{2}{*}{\textit{\textbf{Detector}}}&\textit{\textbf{Channel}}\\
& & &\textit{\textbf{Scheme}}& &\textit{\textbf{Modeling}}\\\hline
III&\multirow{2}{*}{LOS}&\multirow{7}{*}{\makecell{Rotating\\Dipole}}&\multirow{6}{*}{BTS}&\multirow{5}{*}{ED}&Analytical\\\cline{1-1}\cline{6-6}
IV-A and B& & & & &\multirow{3}{*}{4NEC2}\\\cline{1-2}
IV-A and B&\multirow{7}{*}{Scattering}& & & &\\
V-A and B& & & & &\\\cline{1-1}\cline{6-6}
IV-D& & & & &Experiments\\\cline{1-1}\cline{5-6}
V-A and B& & & &\multirow{3}{*}{LSE}&\multirow{2}{*}{4NEC2}\\\cline{1-1}\cline{4-4}
VI-B& & &\multirow{3}{*}{PCS}& &\\\cline{1-1}\cline{3-3}\cline{6-6}
VI-B& &SRR& & &\multirow{2}{*}{Experiments}\\\cline{1-1}\cline{3-3}
VI-B& &XPOL& & &\\\hline
\end{tabular}
\label{tab1}
\end{table}

\section{Analysis in LOS Environments}
\label{sectionIII}
In this section, we consider a LOS propagation environment, i.e. $N^{SC}$ and $N^{RP}$ are equal to zero. Only the source, the tag and the reader are considered. We consider a rotating dipole antenna, a BTS coding scheme, and an ED detector. The carrier frequency $f$ is set to 2.4 GHz. We limit the study to an IPR tag in order to find the best polarization for the tag and some upper bound performance.

\subsection{Selection of the Optimal Polarization}
\label{sectionIIIA}
In this section, we propose an approach to find the orientation of the linear polarization that maximizes $\Delta A$. We assume that the source, the tag and the reader have a perfect linear polarization. Under these assumptions, the direction of the electric field is given by the orientation of the dipole antennas. Let $\Vec{\mathbf{S}}$ be the normalized electric field vector and let $\Vec{\mathbf{T}}$ and $\Vec{\mathbf{R}}$ be the unitary vectors that determine the orientations of the dipoles of the tag and the reader, respectively. For a given position and orientations of the dipole antennas, we approximate the direct source-to-reader signal, $S^{direct}$ (for a normalized electric field $\Vec{\mathbf{S}}$), with the projection of $\Vec{\mathbf{S}}$ over $\Vec{\mathbf{R}}$. The source-to-tag-to-reader signal ($S^{back}$) is approximated with the projection of $\Vec{\mathbf{S}}$ over $\Vec{\mathbf{T}}$, then on $\Vec{\mathbf{R}}$:
\begin{equation}
    S^{direct}=\Vec{\mathbf{S}}\cdot \Vec{\mathbf{R}},
    \label{eq9}
\end{equation}
\begin{equation}
    S^{back}=\big( (\Vec{\mathbf{S}}\cdot \Vec{\mathbf{T}})\Vec{\mathbf{T}}\big)\cdot \Vec{\mathbf{R}},
    \label{eq10}
\end{equation}
where $\cdot$ is the dot product. Then, for a fixed vertical source ($\phi^S=0$) we obtain the following optimal solution:
\begin{equation}
    \begin{cases}
        \theta^T=\theta^R \ [2\pi],\\
        \phi^T=\frac{\phi^R}{2} \ [\frac{\pi}{2}].
    \end{cases}
    \label{eq11}
\end{equation}

From \ref{eq11}, we conclude that there exists an optimal orientation for the tag. Therefore, an IPR tag is expected to outperform an NR tag. The derived optimum orientation of the tag can be interpreted as follows: the best orientation of the IPR tag is obtained when the tag simultaneously maximizes the received signal from the source and maximizes the backscattered signal to the reader. This is obtained when the angle between the source and the tag is equal to the angle between the tag and the reader.

\subsection{4NEC2-based Validation}
\label{sectionIIIB}
We validate the proposed analytical result by using simulations. The 4NEC2 simulation tool takes into account the following elements that are not taken into account in the analytical model: the wire length, the coupling between the elements, the LOS propagation, and the radiation diagram. For a given set of source, tag and reader locations and for a given set of reader orientations, we determine by simulation, through an exhaustive search, the orientation of the tag that maximizes $\Delta A$. The numerical search is performed over a reduced number of angles due to the symmetry: ($\phi^R$,$\theta^R$) in$([0,10, \dots, 90],[0,10, \dots,90])$ and ($\phi^T$,$\theta^T$) in$([0,1, \dots, 90],[0,1, \dots,180])$ degrees. For a given reader orientation, once the optimal tag orientation is found, we compare two unitary vectors giving the best orientation of the tag according to two different methods:
\begin {itemize}
\item ($\Vec{\mathbf{T}}^{best-th}$) that is obtained through the analytical model,
\item ($\Vec{\mathbf{T}}^{best-simu}$) that is obtained through an exhaustive search.
\end {itemize}

\begin{figure}
\centerline{\includegraphics[width=3.5in]{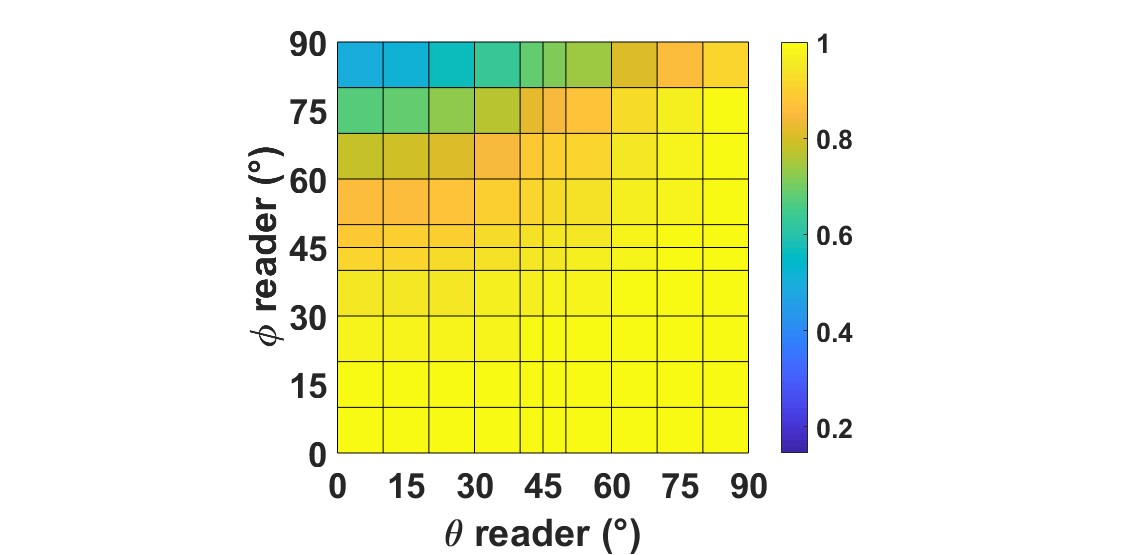}}
\caption{Comparison of the best orientation obtained by using the model and an exhaustive search (dot product of $\Vec{\mathbf{T}}^{best-th}$ and $\Vec{\mathbf{T}}^{best-simu}$). \label{fig3}}
\end{figure}

Fig. \ref{fig3} illustrates the dot product between $\Vec{\mathbf{T}}^{best-th}$ and $\Vec{\mathbf{T}}^{best-simu}$. A high dot product (close to 1) means that the polarization found with the analytical model matches the polarization obtained by exhaustive search. Fig. \ref{fig3} shows that the analytical and simulation results match for more than 80\% of the orientations of the reader, i.e., $\theta^R >50^\circ$ or $\phi^R<45^\circ$. The analytical approach is, therefore, valid in most cases. The cases where the model is not valid correspond to the source being out of the “donut” diagram of the reader dipole and are therefore of low interest.

\subsection{Takeaway Messages from the Analysis}
\label{sectionIIIC}
Based on our analysis, we conclude that, in a LOS propagation environment, the optimal angle of the IPR tag is equal to the mean between two angles: the angle of the source and the angle of the reader. 

\begin{figure}
\centerline{\includegraphics[width=3.5in]{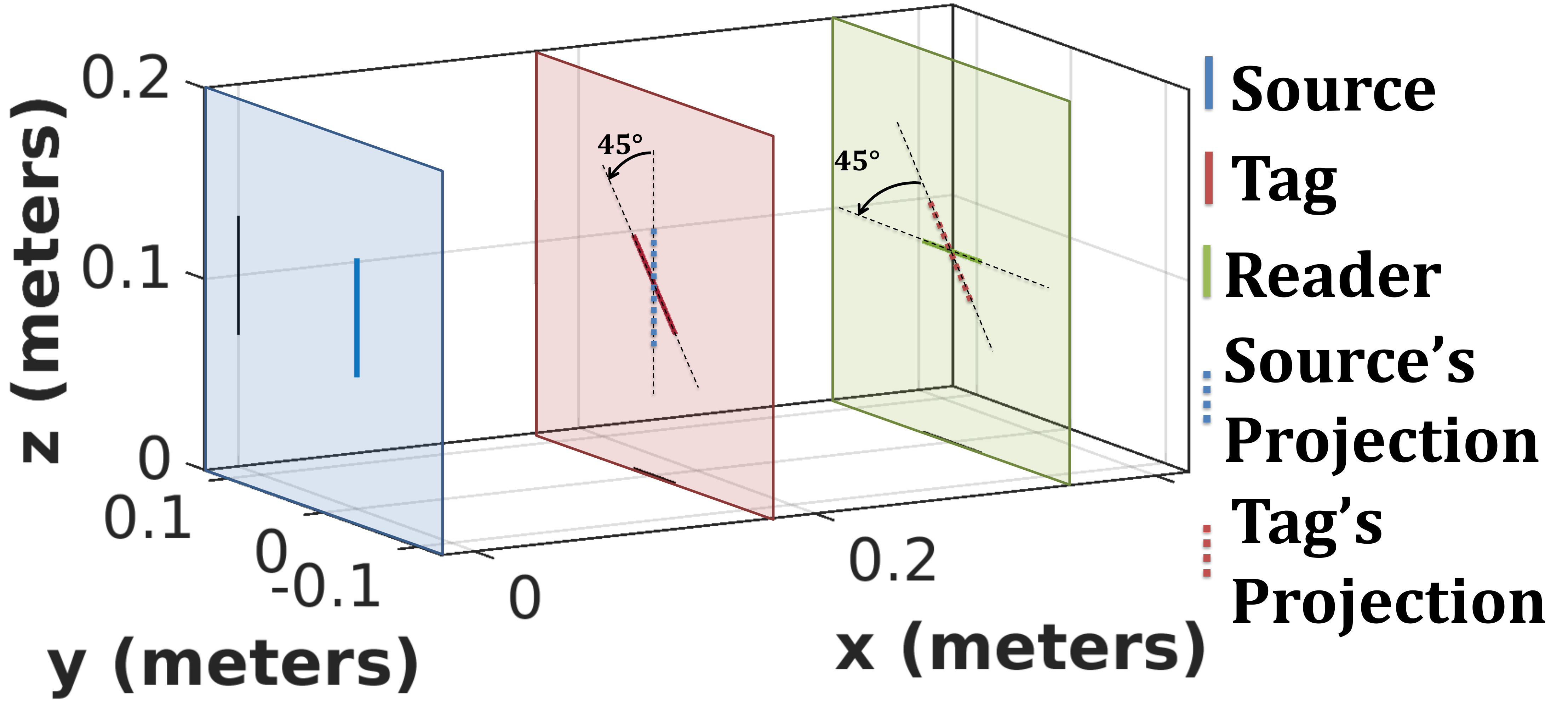}}
\caption{Example of optimum angle for a given source $(\theta^S,\phi^S)=(0^\circ,0^\circ)$ and reader $(\theta^R,\phi^R)=(90^\circ,90^\circ)$ configuration.\label{fig4}}
\end{figure}

The PR tag takes advantage of the mispolarization between the source and the reader. For example, when the source and the reader are orthogonal to each other, the reader can  receive the PR tag signal without being interfered by the direct source-to-reader signal. The current study shows that the IPR tag is better detected by the reader, when it is transmitting its message, with the direction of its linear polarization being at $45^{\circ}$ from the source and $45^{\circ}$ from the reader, as shown in Fig. \ref{fig4}.

In a real environment, in addition to the LOS propagation path studied in this section, additional propagation paths exist (due to scatterers and reflectors). This is studied in the next section.

\section{Analysis in a Rich Scattering Environment}
\label{sectionIV}
In this section, we consider a rich scattering environment, with a large number of scatterers and a few reflective planes. We consider rotating dipole antennas, the BTS coding scheme, and an ED detector.

\begin{table}[ht]
\caption{System model parameters}
\begin{center}
\begin{tabular}{|c|c|c|c|}
\hline
\textbf{\textit{Parameters}} & \textbf{\textit{Details}}& \textbf{\textit{Value}}& \textbf{\textit{Units}} \\
\hline
$(x^S,y^S,z^S)$& Source localization & $(0,0,0.3)$& m \\
\hline
$(\phi^S,\theta^S)$& Source orientation& $(0,0)$& deg \\
\hline
$(x^R,y^R,z^R)$& Reader localization & $(100,0,0.3)$& m \\
\hline
$(\phi^R,\theta^R)$& Reader orientation& $(90,90)$& deg \\
\hline
$(x^T,y^T,z^T)$& Tag localization & $(x^T,y^T,0.3)$& m \\
\hline
$l^{SC}$& Length of scatterers& $\lambda/2$&m \\
\hline
$Z^{R}$& Reader load impedance& $50$&$\Omega$ \\
\hline
$BER^{target }$& Target bit error rate & $10^{-2}$& \\
\hline
\end{tabular}
\label{tab2}
\end{center}
\end{table}

The carrier frequency $f$ is set to 2.4 GHz. We consider the three different types of tags. As mentioned, the 4PR and NR tags have $N_{pol}=4$ and $N_{pol}=1$, respectively. The IPR tag has $N_{pol}=81$. The considered simulation environment accounts for $N^{SC}=20$ scatterers and one ground plane ($N^{RP}=1$). The waves travel from the source to the reader through the LOS path and through multiple non-line-of-sight (NLOS) paths due to presence of the scatterers in proximity of the reader and the ground plane. Whereas the linear polarization of the wave remains unchanged along the LOS path, it may change on the NLOS paths. Hence, at the reader side, the source-to-reader signal is given by the combination of incident waves with distinct linear polarizations. It is expected, however, that the linear polarization of the LOS path is dominant.

\subsection{Numerical Analysis of Polarization and Performance}
\label{sectionIVA}

\begin{figure}
\centerline{\includegraphics[width=3.5in]{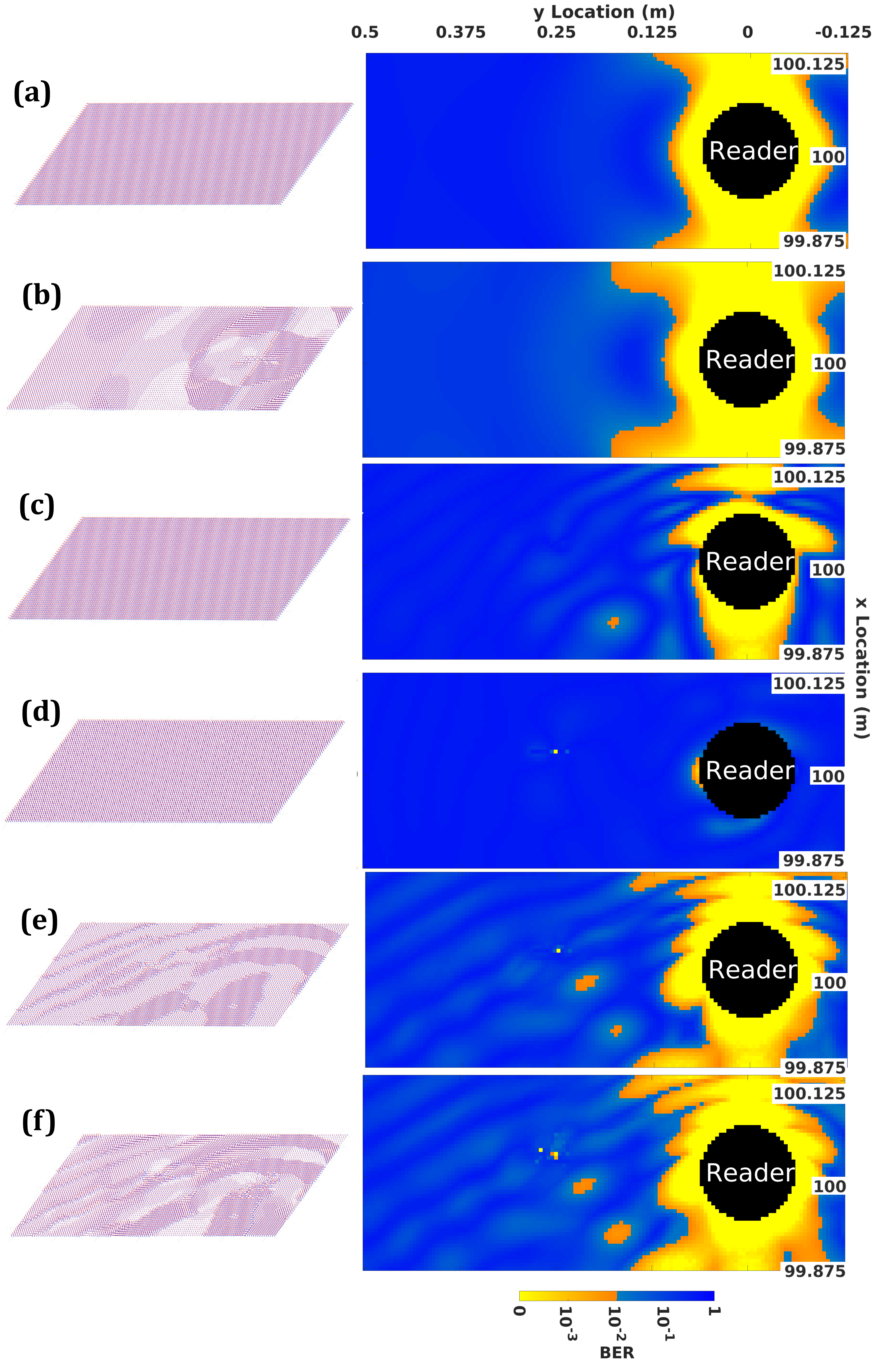}}
\caption{$BER$ maps showing the tag location that guarantees QoS (in orange) for the configurations : (a) LOS-NR-Best, (b) LOS-IPR, (c) SCAT-NR-Best, (d) SCAT-NR-Worst, (e) SCAT-4PR, (f) SCAT-IPR. On the left of each subfigure, the tag orientation map (i.e. the velvet carpet) is illustrated.\label{fig5}}
\end{figure}

In this section, we illustrate the best linear polarization of the tag by using 4NEC2 simulations. Source and reader orientations are fixed and orthogonal to each other, as it is a worst case scenario in terms of received SNR. We analyze the performance of the communication link as a function of the tag coordinates ($x^T$,$y^T$), i.e., we illustrate 2D spatial maps of the BER. Other simulation parameters are detailed in Table \ref{tab2}. We study the three types of tag for different configurations of their orientation and for different environments (LOS or with scattering). Fig. \ref{fig5} illustrates the BER maps of some of the configurations in Table \ref{tab3} by assuming $SNR^{Tx}=116 \ dB$. The spatial maps of the BER show the locations where the tag can be detected by the reader with the target QoS ($BER<BER^{target}$). Light colors (yellow or red) indicate where the QoS can be achieved and dark colors (blue) indicate locations where the QoS cannot be achieved. Each subfigure of Fig. \ref{fig5} is accompanied by a “velvet carpet” illustrating the tag’s best orientation depending on the tag’s location. Each position on the velvet carpet corresponds to a tested position of the tag. The thread of the carpet at a given position illustrates the best orientation of the tag for the considered position. The velvet carpet illustrations in Fig. \ref{fig5}-a,c,d show that the orientation is uniform for NR tags and Fig. \ref{fig5}-b,e,f show that the orientation is non-uniform for PR tags.

\begin{table}[ht]
\caption{Tag and environment configurations}
\begin{center}
\begin{tabular}{|c|c|c|}
\hline
 &\textit{\textbf{LOS}} &\textit{\textbf{Scattering}} \\
\hline
NR tag in the worst orientation & LOS-NR-Worst & SCAT-NR-Worst \\
\hline
NR tag in the best orientation & LOS-NR-Best & SCAT-NR-Best \\
\hline
4PR tag & LOS-4PR & SCAT-4PR \\
\hline
IPR tag& LOS-IPR & SCAT-IPR \\
\hline
\end{tabular}
\label{tab3}
\end{center}
\end{table}

In Fig. \ref{fig5}-a and in Fig. \ref{fig5}-b, we show the results obtained for the configuration in a LOS environment. Fig. \ref{fig5}-a illustrates the BER for the optimum orientation in LOS: $\theta^T=90^\circ$ and $\phi^T=45^\circ$. Fig. \ref{fig5}-b shows the contrast map in a LOS environment for the IPR tag. We observe that a PR tag has limited effects in a LOS channel if the source and the reader have a cross-polarization. 
The following rotating dipole antenna tags are compared in a scattering environment.
\begin{itemize}
    \item 	\textbf{NR Tag}: We study the performance of the NR tag when its fixed orientation is, on average, the optimal one. We determine numerically this best orientation. We observe, that even in a scattering environment (Fig. \ref{fig5}-c), we obtain the same best orientation as in a LOS environment. Indeed, the optimum orientation in LOS, obtained by applying \ref{eq7} to the parameters listed in Table \ref{tab2}, is $\theta^T=90^\circ$ and $\phi^T=45^\circ$. This is due to the fact that our studied scattering environment is close to a LOS environment. For comparison, Fig. \ref{fig5}-d illustrates the performance of the NR tag when its fixed orientation is, on average, the worst ($\theta^T=90^\circ$ and $\phi^T=90^\circ$), i.e., for which a very small amount of backscattered signal can be detected by the reader.
    
    \item \textbf{4PR tag}: The compact reconfigurable antennas presented in \cite{ref8} have four patterns with distinct dominant linear polarization directions. The 4PR tag is a simplified model of such existing antennas. The angles of the 4 polarizations are set to $(x^T,y^T)={(0,90),(45,90),(90,90),(135,90)}$ and they correspond to the main polarization directions of one of the antennas from \cite{ref8}. The SNR contrast maps are computed for the 4 polarizations of the tag.
    
    \item \textbf{IPR tag}: We consider an IPR tag with $N_{pol}=81$. The orientations of the IPR tag $(x^T,y^T)$ are uniformly distributed in $([0,180],[0,180])$ degrees. We compute the BER maps for each of $N_{pol}$ polarizations.
    
\end{itemize}

As for the IPR and 4PR tags, we determine the polarization of the tag that minimizes the BER for every location of the tag based on the $N_{pol}$ computed maps of each tag. We obtain the corresponding optimal BER maps as illustrated in the Fig. \ref{fig5}-e,f. As expected, Fig. \ref{fig5} shows that the IPR tag outperforms all types of tags. We also observe that the more realistic 4PR tag, with 20 times fewer available polarizations than the IPR tag, offers performance that is close to the IPR tag. The 4PR tag outperforms the NR tag, even if the NR tag employs the optimal orientation of the LOS system. Compared to the NR tag, the 4PR tag is more robust to the impact of scatterers and polarization mismatch. 

In the next sub-section, we study the performance of the tags as a function of $SNR^{Tx}$.

\subsection{Outage Probability Analysis}
\label{sectionIVB}

In the previous section, we have shown that a PR tag can improve the performance of an AmB system in terms of BER. In this section, we numerically evaluate the outage probability for each configuration listed in Table \ref{tab3}. The outage probability is computed over a target coverage area, i.e., over tag’s locations (in meters) defined by: $x^T=x^R+\alpha\times0.001$ and $y^T=y^R+\beta\times0.001$, with $0.5\lambda<D^{(T-R)}<3\lambda$ and $\alpha,\beta,\in Z$, where $D^{(T-R)}$ corresponds to the Euclidian distance between the tag and the reader. The BER is calculated as a function of $SNR^{Tx}$, for every tag location and orientation and for a given configuration of the source, the reader, the scatterers and the reflective planes. We compute the outage probability, based on \ref{eq7}, for the three types of tags.

\begin{figure}
\centerline{\includegraphics[width=3.5in]{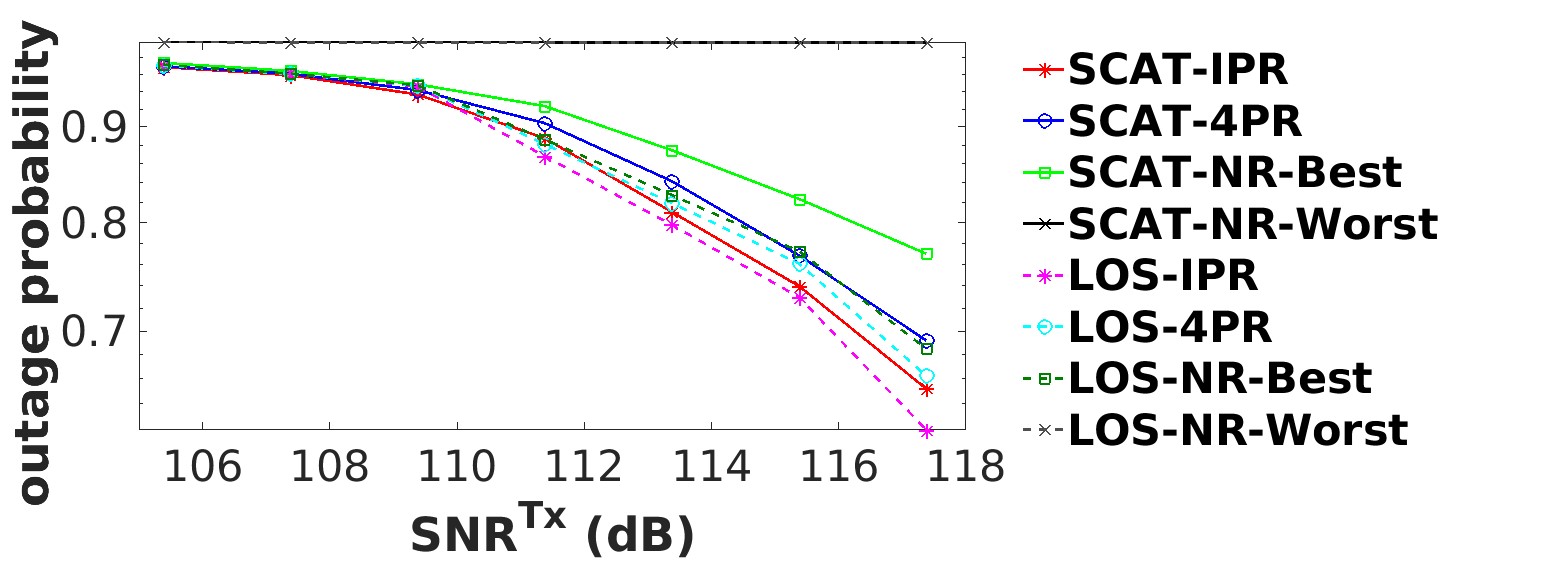}}
\caption{Outage probability simulated as a function of $SNR^{Tx}$.\label{fig6}}
\end{figure}

\begin{figure}
\centerline{\includegraphics[width=3.5in]{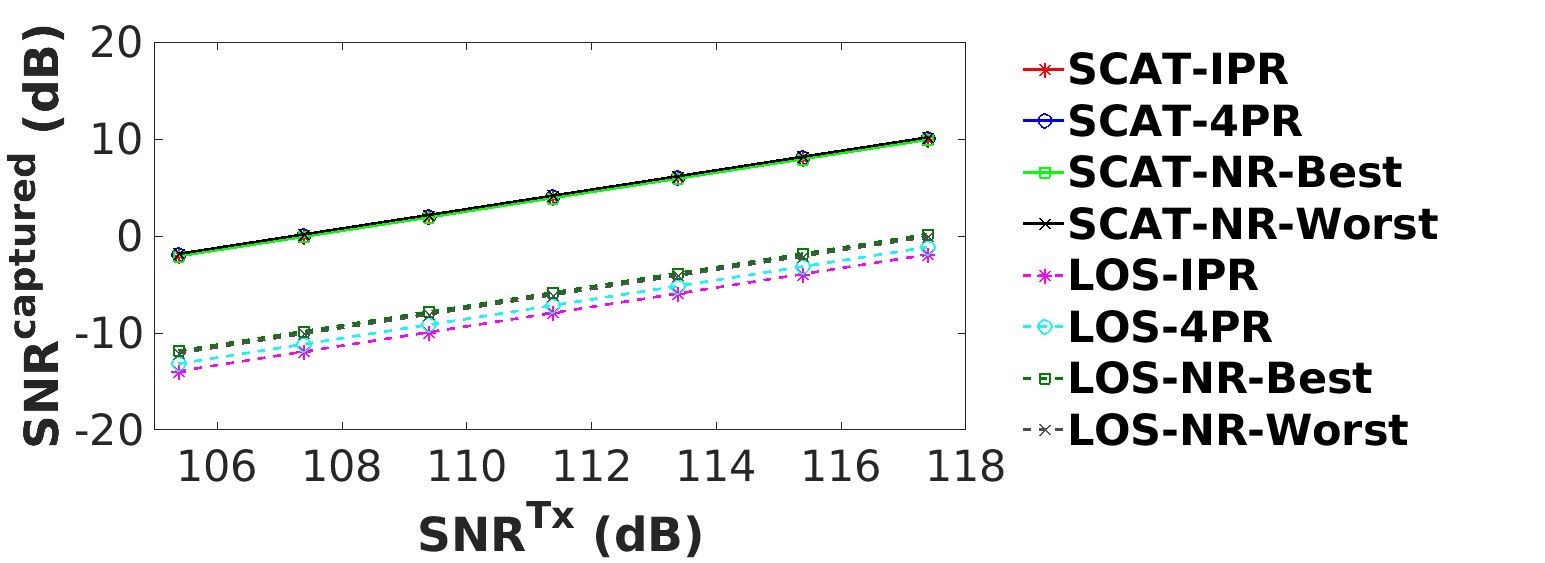}}
\caption{$SNR^{captured}$ by the reader as function of the $SNR^{Tx}$.\label{fig7}}
\end{figure}

Fig. \ref{fig6} shows that the IPR tag attains the best performance. It provides the lower bound performance for this environment configuration. The NR tag, even configured to operate at the optimal orientation, has a high outage probability, since it is not robust to scattering. Fig. \ref{fig7} shows the average captured SNR for each configuration as a function of $SNR^{Tx}$. In LOS, the amount of received power corresponds to the power backscattered by the tag, which depends on the tag polarization. However, as expected, in a scattering environment, the $SNR^{captured}$ does not depend on the tag polarization (as it is measured when the tag is transparent) and only depends on the scattering. In the LOS-NR-Worst configuration, the NR tag and the reader are orthogonal to the source, thus the reader receives close-to-zero signals from the source and from the tag. We observe that increasing the number of reconfigurable polarizations $N_{pol}$ of the tag improves the system performance. In addition, this improvement is boosted by the presence of scatterers. Finally, we observe that, even with a limited number of polarization orientations ($N_{pol}=4$), the performance of the 4PR tag is close to the IPR tag.

\subsection{Impact of the scatterers}
\label{sectionIVC}
\begin{figure}
\centerline{\includegraphics[width=3.5in]{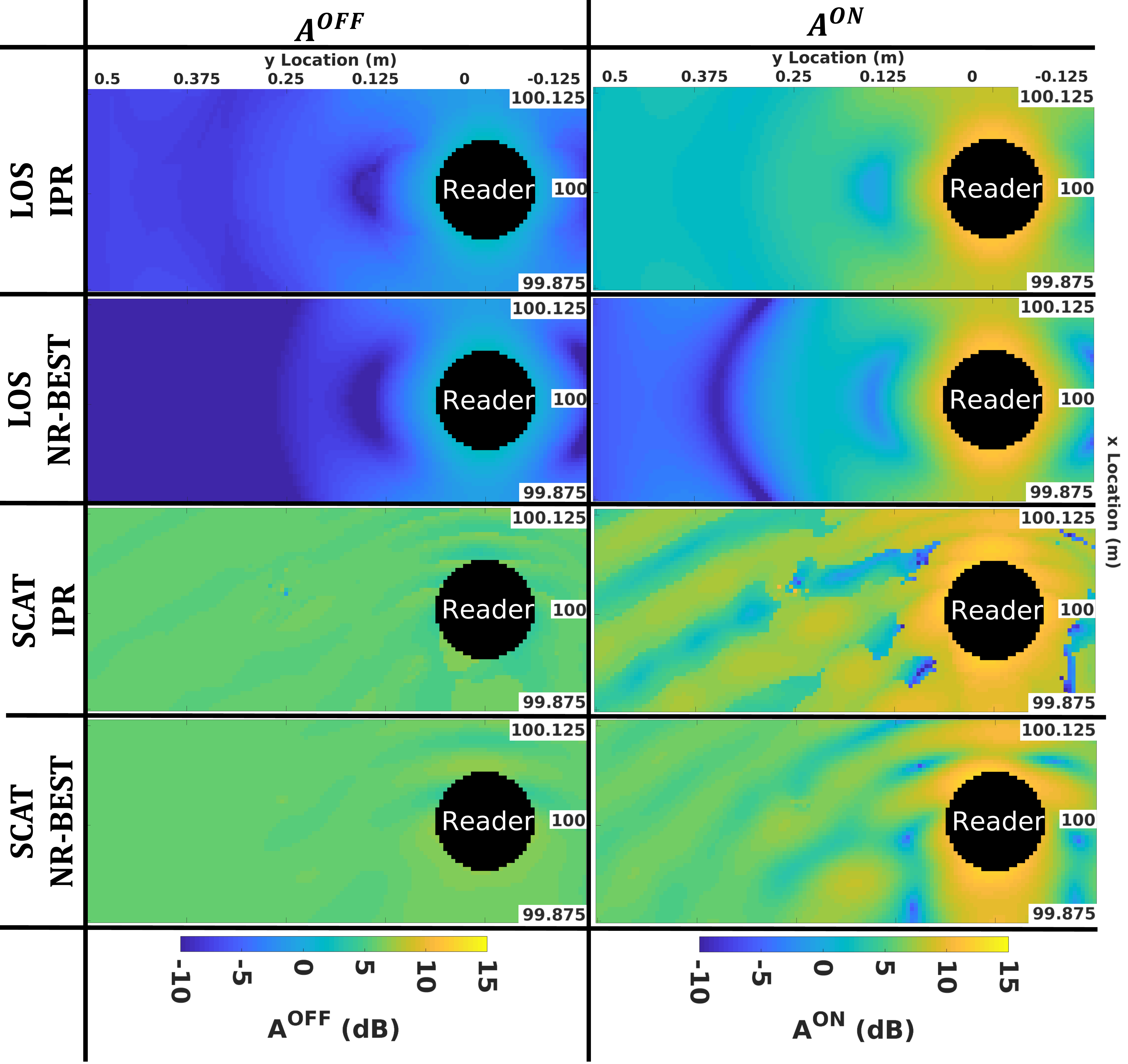}}
\caption{Amplitude maps for the transparent and backscattering state of the IPR and NR tags in scattering and LOS environment.\label{fig8}}
\end{figure}

Fig. \ref{fig8} illustrates the amplitude of the signal $A^{OFF}$, respectively $A^{ON}$, received by the reader when the tag is transparent, respectively backscattering, as a function of the position of the tag in space, for two different propagation environments (LOS and with scattering), for two different tags (NR-Best and IPR), and for the ED detector. As the presence of scatterers results in additional paths in the source-to-reader propagation channel (in addition to the LOS path), the amplitude of the signal $A^{OFF}$, when the tag is in the transparent state, is much higher in scattering condition compared to the received signal $A^{OFF}$ in LOS condition. As a consequence, in the scattering condition, the additional path due to the tag being in backscattering state is much less visible by the reader, compared to the LOS condition.

\subsection{Experimental Verification}
\label{sectionIVD}

\begin{figure}
\centerline{\includegraphics[width=3.5in]{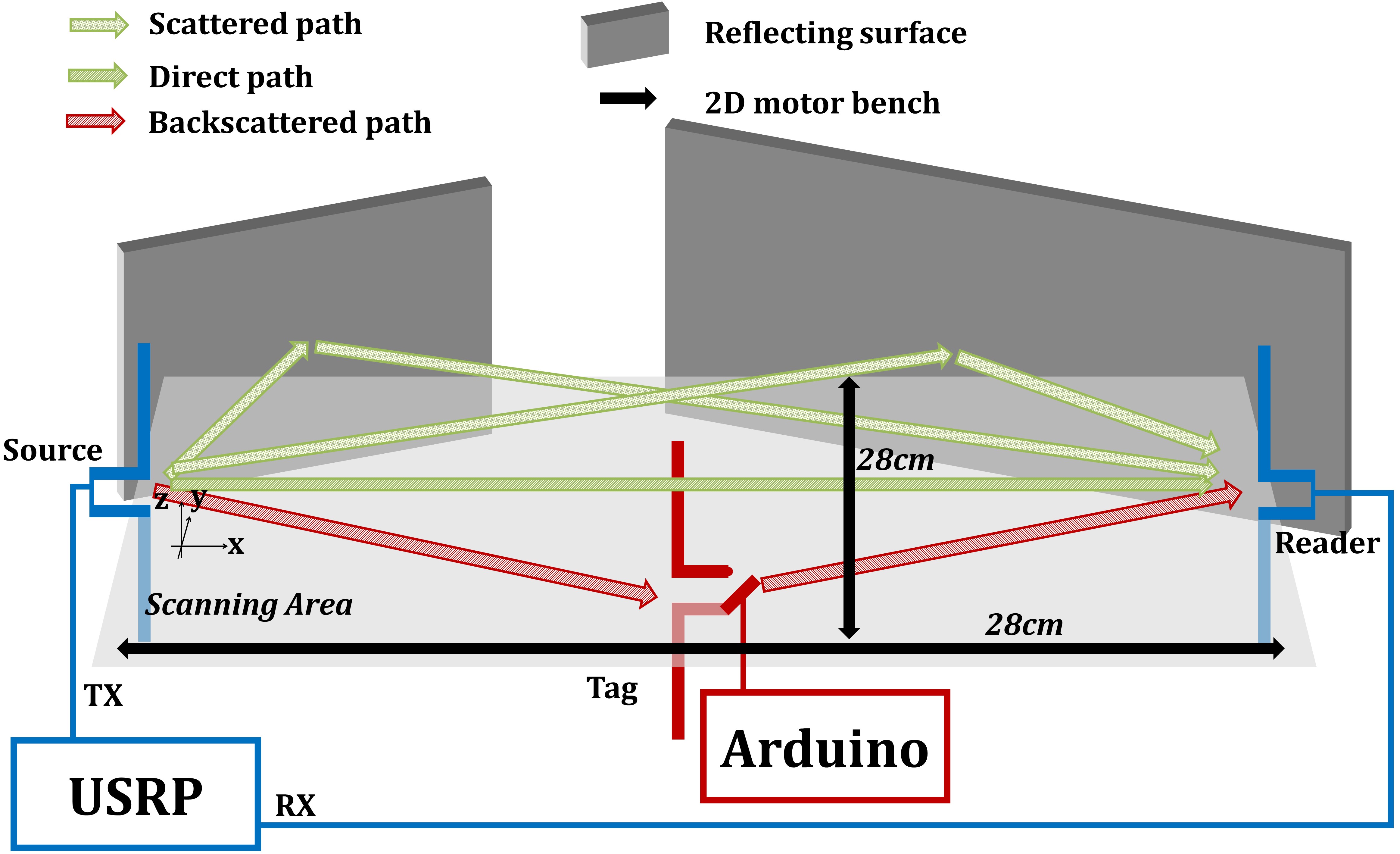}}
\caption{Experimental setup.\label{fig9}}
\end{figure}

\begin{figure}
\centerline{\includegraphics[width=3.5in]{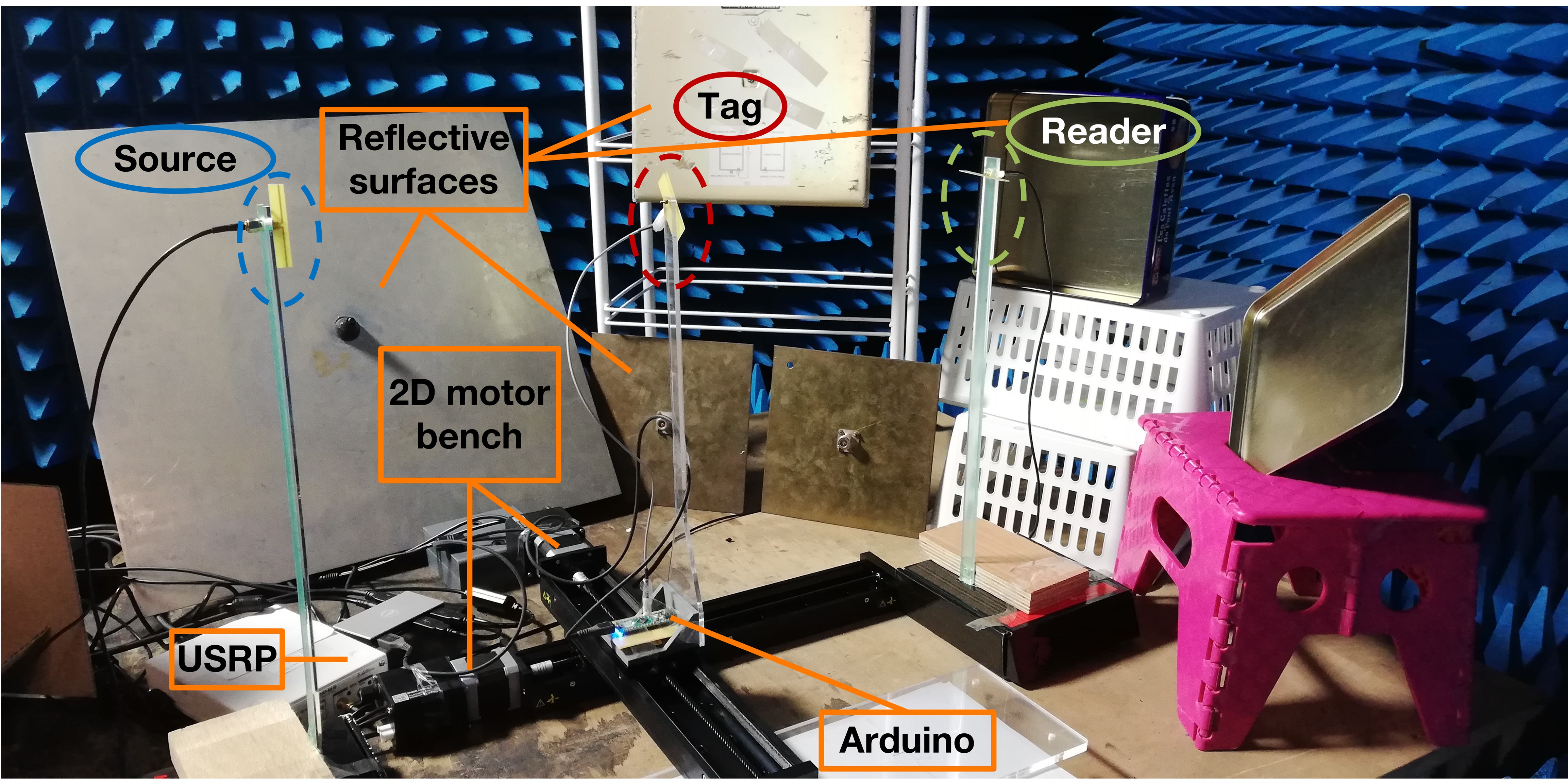}}
\caption{Photo of the experimental setup.\label{fig10}}
\end{figure}

In the previous section, we have shown, with the aid of simulations, that a PR tag outperforms an NR tag. In this section, we validate this observation through experiments conducted in a semi-anechoic chamber. In our experimental setup, we reproduced, as closely as possible, the conditions modeled in the simulation environment. Instead of scatterers made of conductive lines, we deploy reflective planes in the environment ($N^{SC}=0, N^{SC}=6$). Each of them has a different surface area and is placed randomly around the system, with different orientations and locations, as illustrated in Fig. \ref{fig9} and Fig. \ref{fig10}. The source is located at a shorter distance from the tag and the reader (0.35m) as compared with the simulation results (100m). This is necessary because of the limited size of the semi-anechoic chamber. To measure the signal contrast maps experimentally, the source and the reader are installed at fixed locations. The mechanically rotating dipole antenna of the tag is mounted on two motorized rails whose length is 0.3m.

Each element is composed of a dipole antenna made of two strands with a total length of 6.25 cm. We connect the source and the reader to two different ports of a USRP B210. We use a GNU Radio to control the USRP and process the source and received signals. The tag’s antenna is connected to an Arduino platform that controls the impedance connected to the two strands using a PIN diode.

We place the source at the origin of the axis, $(x^S, y^S, z^S)=(0, 0, 0)$, with a vertical orientation, $(\phi^S, \theta^S)=(0, 0)$. The reader is placed at position $(x^R, y^R, z^R)=(0.35, 0, 0)$ and is in cross-polarization with the source $(\phi^R, \theta^R)=(90, 90)$. The tag is moved along a linear trajectory perpendicular to the line connecting the source and the reader, such that $x^T=xmin+nx\times step$ and $y^T=ymin+ny\times step$. 
We define $xmin=0.03m$ and $ymin=-0.15m$ so that the tag scans the area between the source and the reader, given the limited range of $0.28m$.
The step of the tag displacement is $step=10^{-2}m$ along the x and y axis and $nx,ny \in [0,1 \dots, 28]$.

\begin{figure}
\centerline{\includegraphics[width=3.5in]{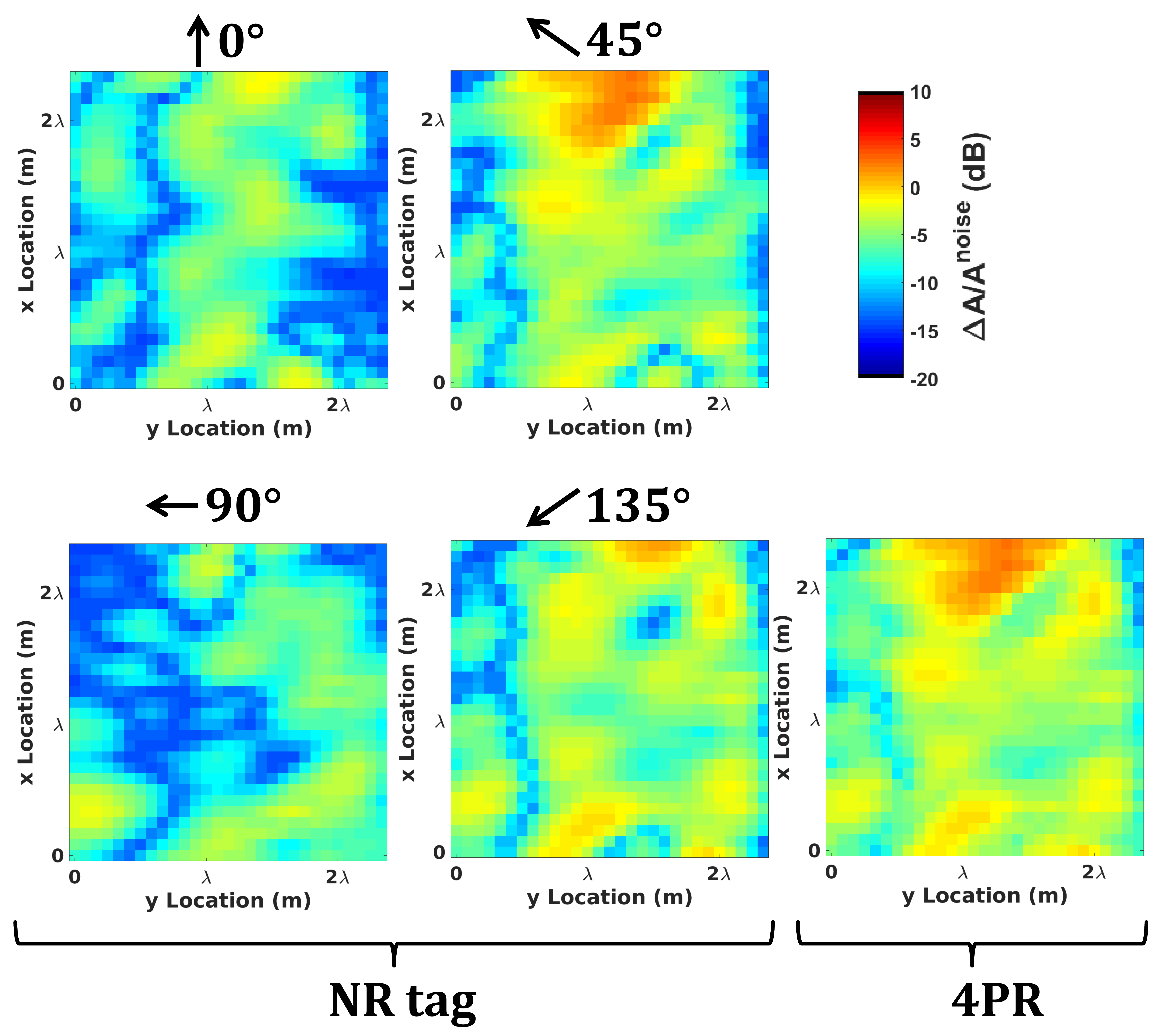}}
\caption{Experimental maps for four fixed orientations of the NR dipole tag $(\phi^T=0^\circ,45^\circ,90^\circ,135^\circ,\phi^T=90^\circ)$ and the map for the 4PR tag: the optimum signal contrast map obtained after selection of the best polarization among the 4 available ones.\label{fig11}}
\end{figure}

We acquire the maps for the signal amplitude of the backscattering and transparent states, $A^{ON}$, $A^{OFF}$, respectively, the noise signal amplitude $A^{noise}$, and then calculate the $\Delta A/A^{noise}$ maps. We acquire the signal contrast maps for each angle of the 4PR tag. We obtain 4 maps for the signal contrast as shown in Fig. \ref{fig11}.

We process these maps to determine the map with the best signal contrast. Experimental results show that selecting among four polarization orientations improves the performance of an AmB system (Fig. \ref{fig11}). Even though the experiment is not a perfect replica of the simulations, experimental and simulation results are consistent: both show that, in a complex environment, the maps depend on the orientation of the polarization of the tag, and that a PR tag is more robust.

\subsection{Takeaway Message from the Analysis}
\label{sectionIVE}
Based on our study based on simulations and experiments, we conclude that the use of a PR tag improves the performance of an AmB system in a rich scattering environment. Based on our study, a simple 4PR tag provides good performance similar to an ideal tag.

\section{Impact of the Reader Detection Scheme}
\label{sectionV}
In this section, we compare the performance of the ED and LSE detectors. We consider rotating dipole antennas, the BTS coding scheme, and a scattering environment. The configuration of the system is given in Table \ref{tab3}. We evaluate the BER maps for locations of the tag around the reader. The three dipole tags IPR, 4PR and NR are considered. The map is shown for a fixed $SNR^{captured}=12 dB$, which is equivalent to $SNR^{Tx}=116 dB$ in the scattering configuration.

\subsection{Visualization of Polarization and Performance}
\label{sectionVA}
\begin{figure}
\centerline{\includegraphics[width=3.5in]{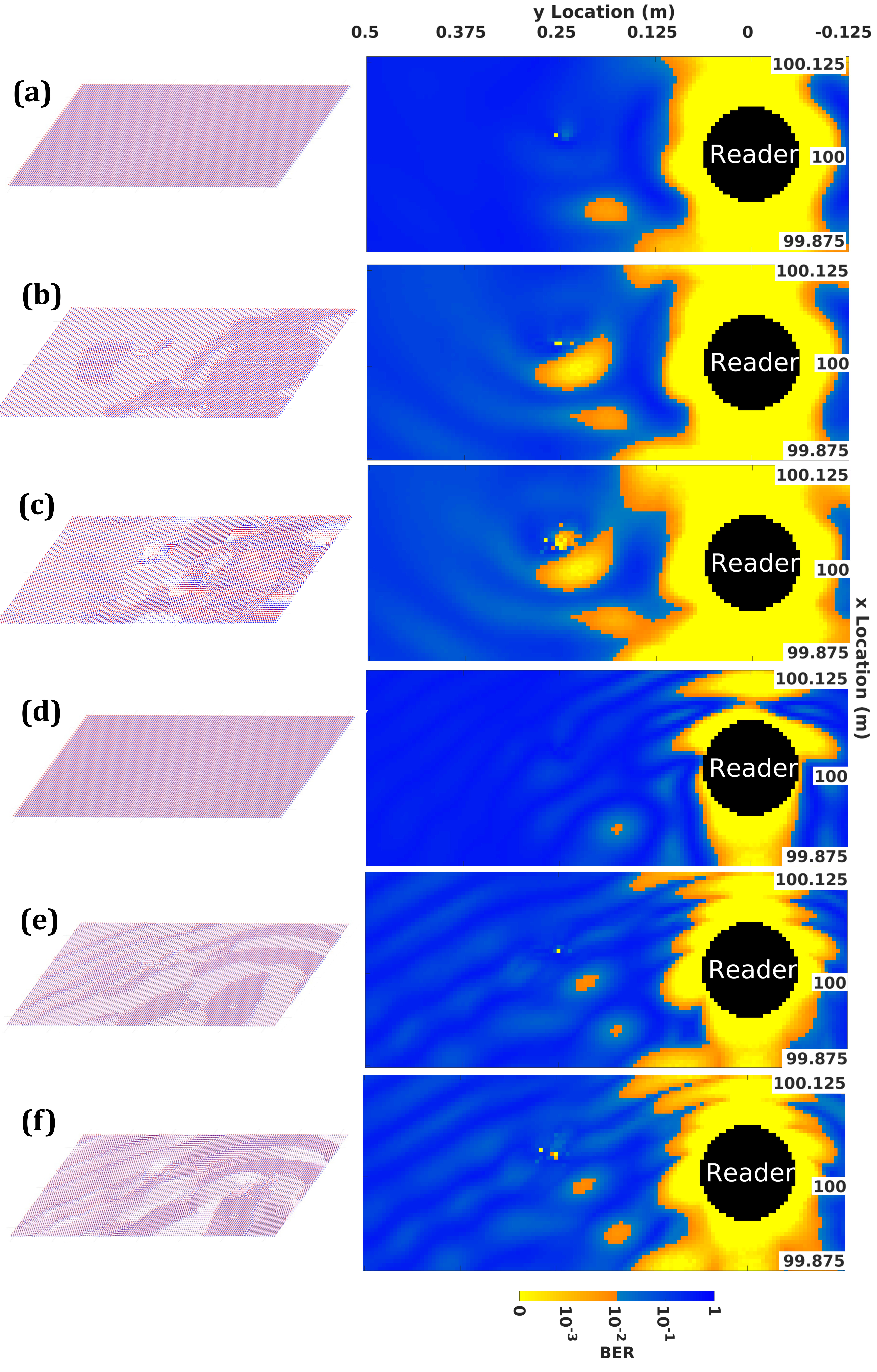}}
\caption{$BER$ maps showing the tag location that guarantees QoS (in orange) for the configurations: (a) LSE-NR-Best, (b) LSE-4PR, (c) LSE-IPR, (d) ED-NR-Best, (e) ED-4PR, (f) ED-IPR. On the left of each subfigure, the tag orientation map is illustrated.\label{fig12}}
\end{figure}

Fig. \ref{fig12} shows the BER map for the three tags and for the two detectors. Similar to the previous section, the IPR tag provides the best performance for both detectors. The LSE detector, however, yields better BER performance and a larger coverage area. Even the NR tag that uses the best polarization orientation and the LSE detector provides better performance than an IPR tag that uses an ED detector. The LSE detector has better performance than  an ED as it compares the amplitude and the phase of the received signal with the estimated channel. 

\subsection{Outage Probability Analysis}
\label{sectionVB}
In this section, we numerically evaluate the outage probability for each configuration setup reported in Table \ref{tab3}. The outage probability is computed over a target coverage area, defined such that the Euclidean distance $D^{(T-R)}$ between the tag and the reader lies in the range  $0.5\lambda<D^{(T-R)}<3\lambda$. The $BER$ is calculated as a function of $SNR^{captured}$ for every tag location and orientation and for a given configuration of the source, the reader, the scatterers and the reflective planes.
\begin{figure}
\centerline{\includegraphics[width=3.5in]{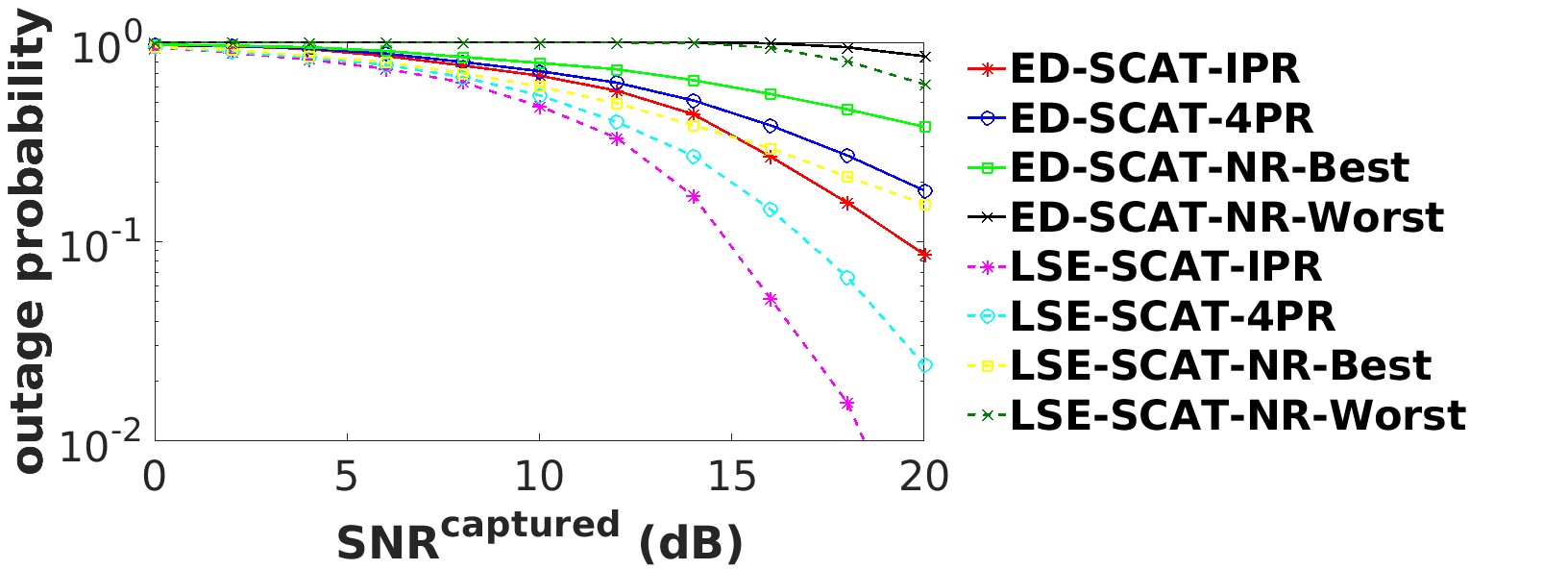}}
\caption{$BER$ simulated for the ED and LSE detectors as a function of the SNR.\label{fig13}}
\end{figure}  
In \ref{fig13}, we observe that the LSE detector improves the performance compared with the ED detector. The IPR tag combined with the LSE detector outperforms the IPR tag combined with the ED detector. The 4PR has good performance that is close to the IPR tag. In conclusion, the LSE detector outperforms the ED, and the 4PR tag with LSE provides close to ideal performance.

\section{Impact of the Antenna Radiation Pattern}
\label{sectionVI}
In this section we compare the performance of the SSR antenna, the XPOL antenna, and the 4PR rotating dipole antenna. We consider The PCS coding scheme and the LSE detector.

\subsection{Experimental Setup}
\label{sectionVIA}
\begin{table}[ht]
\caption{Antenna Configurations}
\setlength{\tabcolsep}{3pt}
\begin{center}

\begin{tabular}{|p{53pt}|p{54pt}|p{54pt}|p{54pt}|}
\hline
\textit{\textbf{Antenna}}&\textbf{\textit{Coding}} \par \textit{\textbf{Scheme}}&\textit{\textbf{Channel}} \par \textit{\textbf{Modelling}}&\textit{\textbf{Number}} \par \textit{\textbf{of receiving}} \par \textit{\textbf{antenna}} \\
\hline
SRR&PCS&Experimental&3\\
\hline
XPOL&PCS&Experimental&3\\
\hline
Dipole 4PR&PCS&Simulation&1\\
\hline
\end{tabular}
\label{tab4}
\end{center}
\end{table}
The configuration setups studied in this section are reported in Table \ref{tab4}. The propagation channel is extracted from measurements conducted in a reverberation chamber. The reverberation chamber has the advantage to provide a time invariant environment and multipath.

\begin{figure}
\centerline{\includegraphics[width=3.5in]{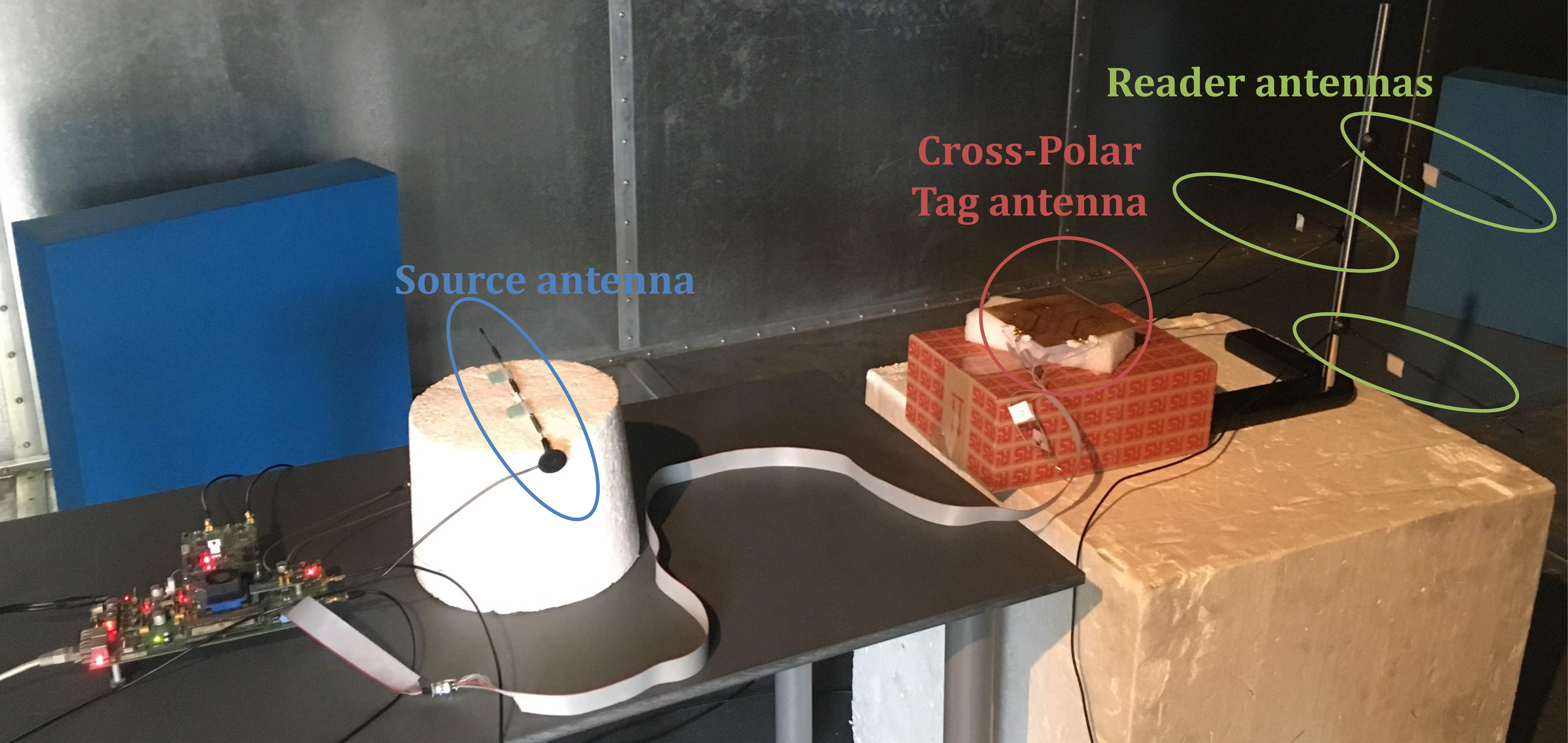}}
\caption{Experimental setup for the XPOL antenna.\label{fig14}}
\end{figure}

\begin{figure}
\centerline{\includegraphics[width=3.5in]{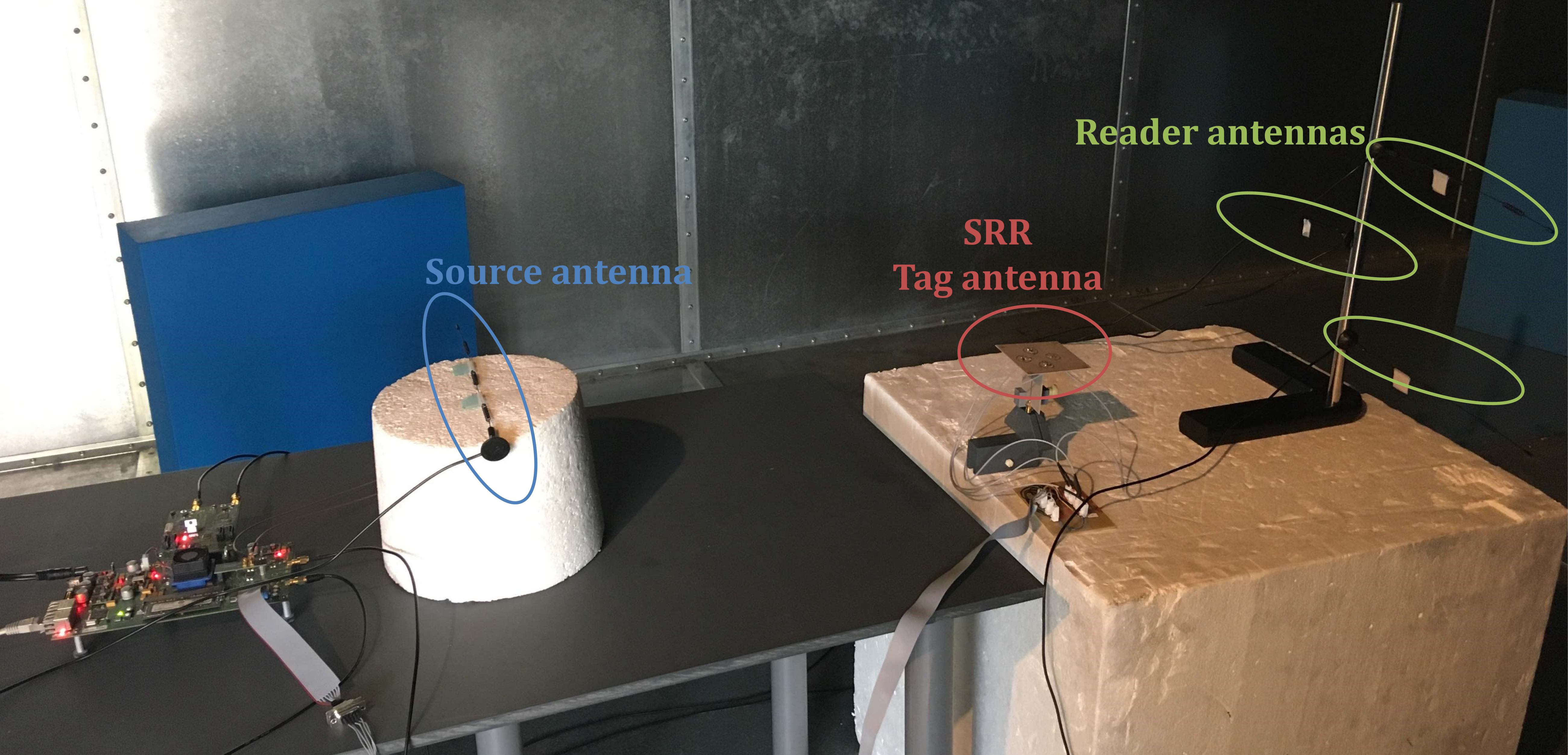}}
\caption{Experimental setup for the SRR antenna.\label{fig15}}
\end{figure}

\begin{figure}
\centerline{\includegraphics[width=3.5in]{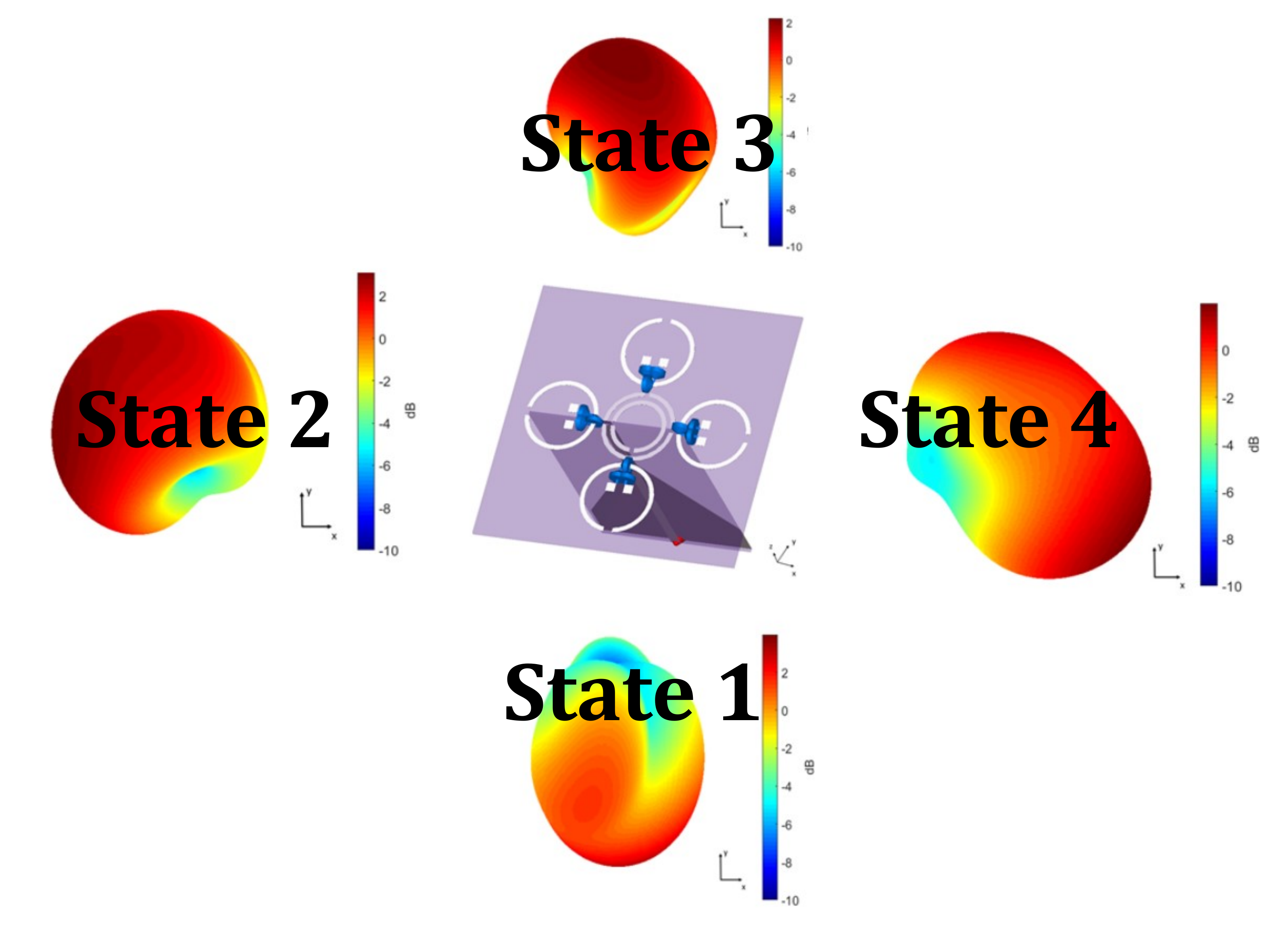}}
\caption{SRR polarization patterns according to the state number.\label{fig16}}
\end{figure}

\begin{figure}
\centerline{\includegraphics[width=3.5in]{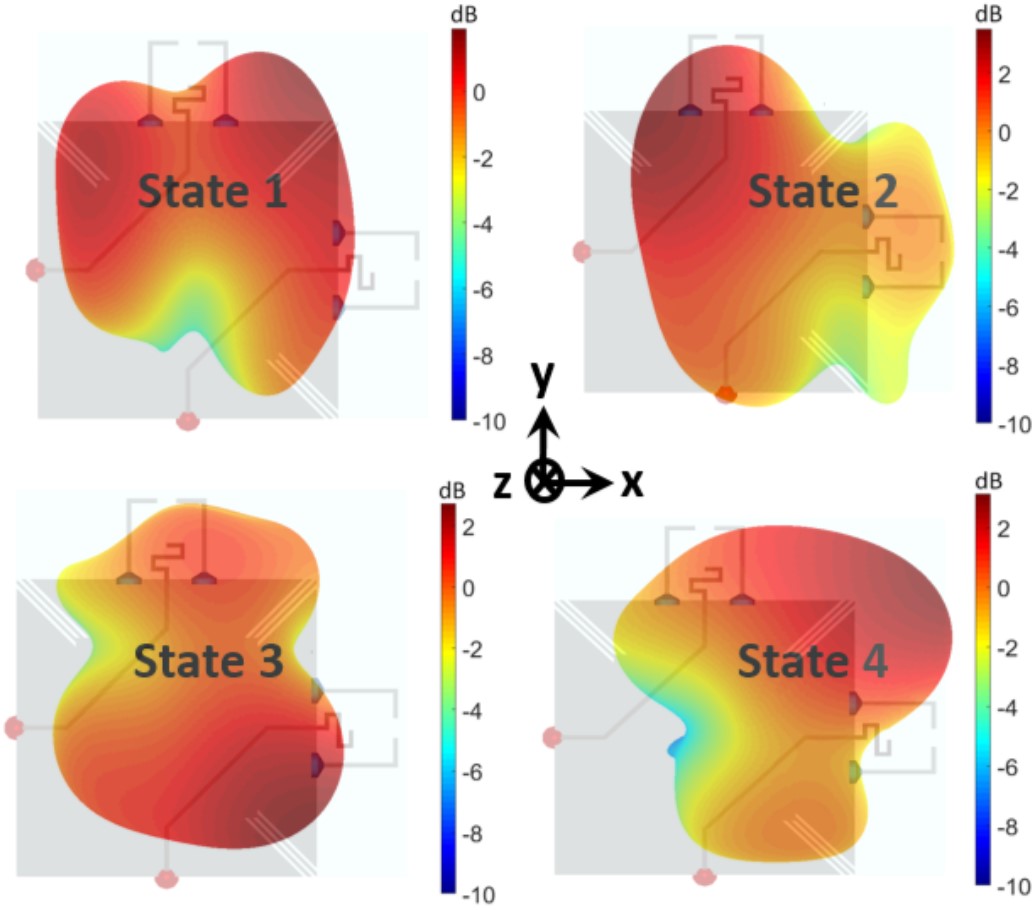}}
\caption{XPOL polarization patterns according to the state number.\label{fig17}}
\end{figure}

\begin{figure}
\centerline{\includegraphics[width=3.5in]{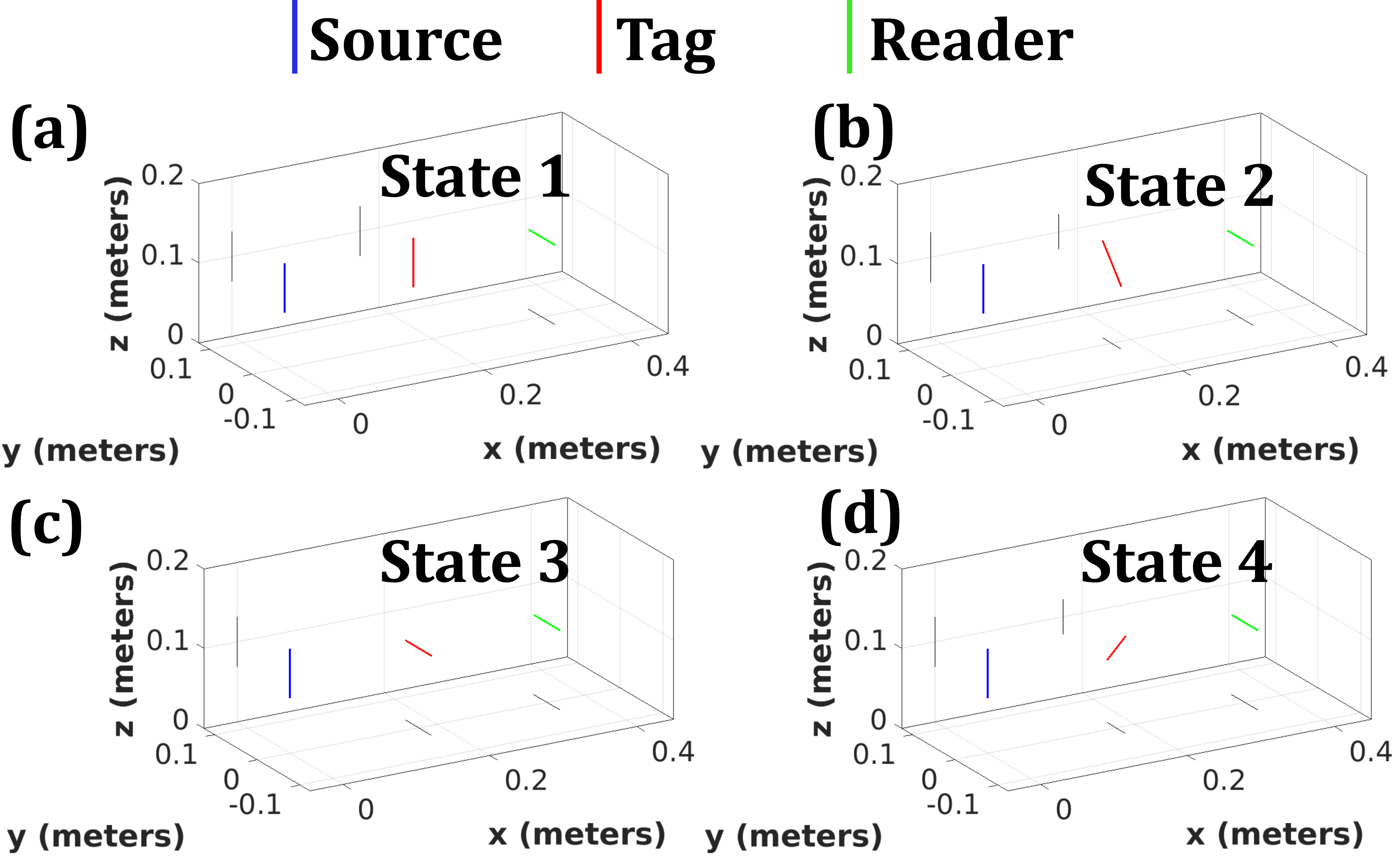}}
\caption{4PR polarization orientation according to the state $(\phi^T=0^\circ,45^\circ,90^\circ,135^\circ,\theta^T=90^\circ)$.\label{fig18}}
\end{figure}

Fig. \ref{fig14} and Fig. \ref{fig15} illustrate the experimental setup of the system for the two compact reconfigurable antennas. For each antenna we consider 4 polarization patterns as depicted in Figs. \ref{fig16}, \ref{fig17}, \ref{fig18}. Based on the PCS coding scheme, the tag switches between two polarization patterns in order to transmit the bits 1 and 0. We evaluate the performance depending on the two chosen patterns for each antenna. 

\begin{table}[ht]
\caption{Configuration of the PCS Based on the State of the Antenna}
\setlength{\tabcolsep}{3pt}
\begin{tabular}{|p{47pt}|p{27pt}|p{27pt}|p{27pt}|p{27pt}|p{27pt}|p{27pt}|}
\hline
\textit{\textbf{Configuration}}&\textit{\textbf{2:1}}&\textit{\textbf{3:1}}&\textbf{\textit{4:1}}&\textit{\textbf{3:2}}&\textbf{\textit{4:2}}&\textit{\textbf{4:3}}\\
\hline
Bit 1&State 2&State 3&State 4&State 3&State 4&State 4\\
\hline
Bit 0&State 1&State 1&State 1&State 2&State 2&State 3\\
\hline
\end{tabular}
\label{tab5}
\end{table}

\subsection{Results}
\label{sectionVIB}
\begin{figure}
\centerline{\includegraphics[width=3.5in]{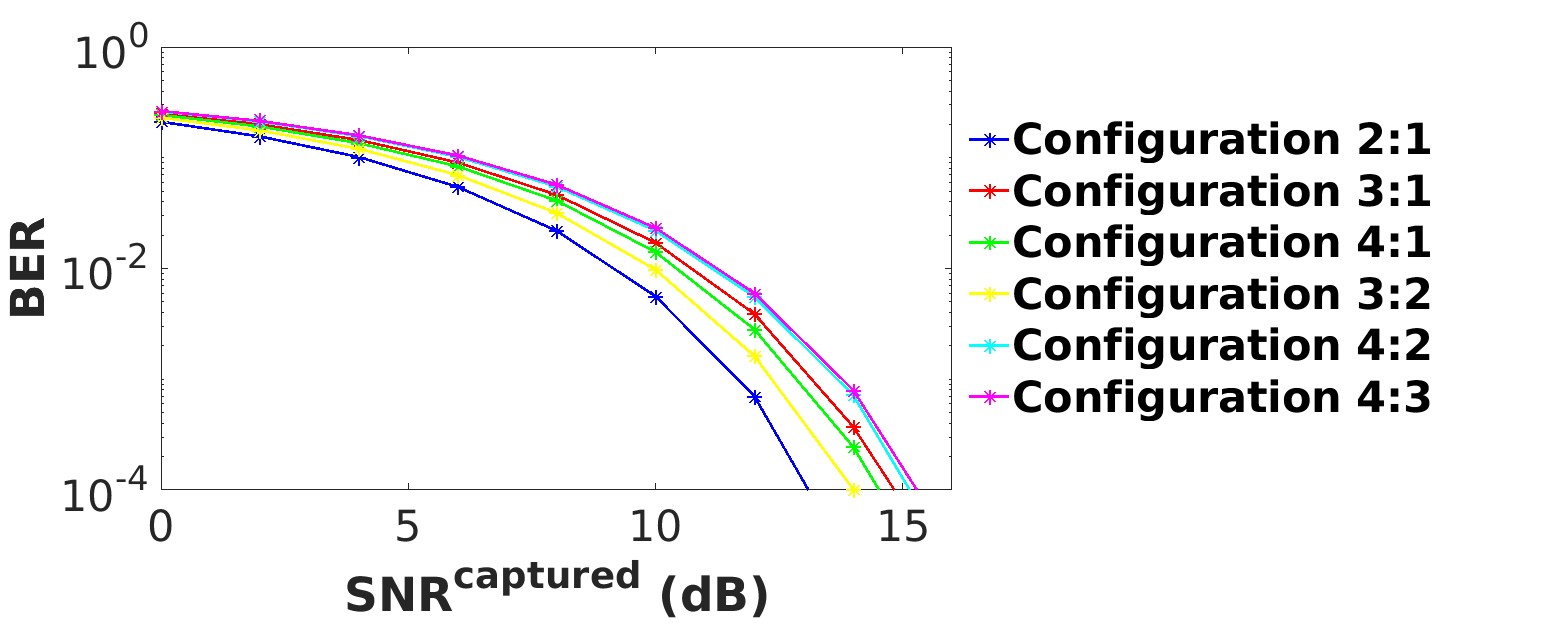}}
\caption{Fig. 18.	BER rate for a the different configurations of the PCS of the SRR antenna.\label{fig19}}
\end{figure}

\begin{figure}
\centerline{\includegraphics[width=3.5in]{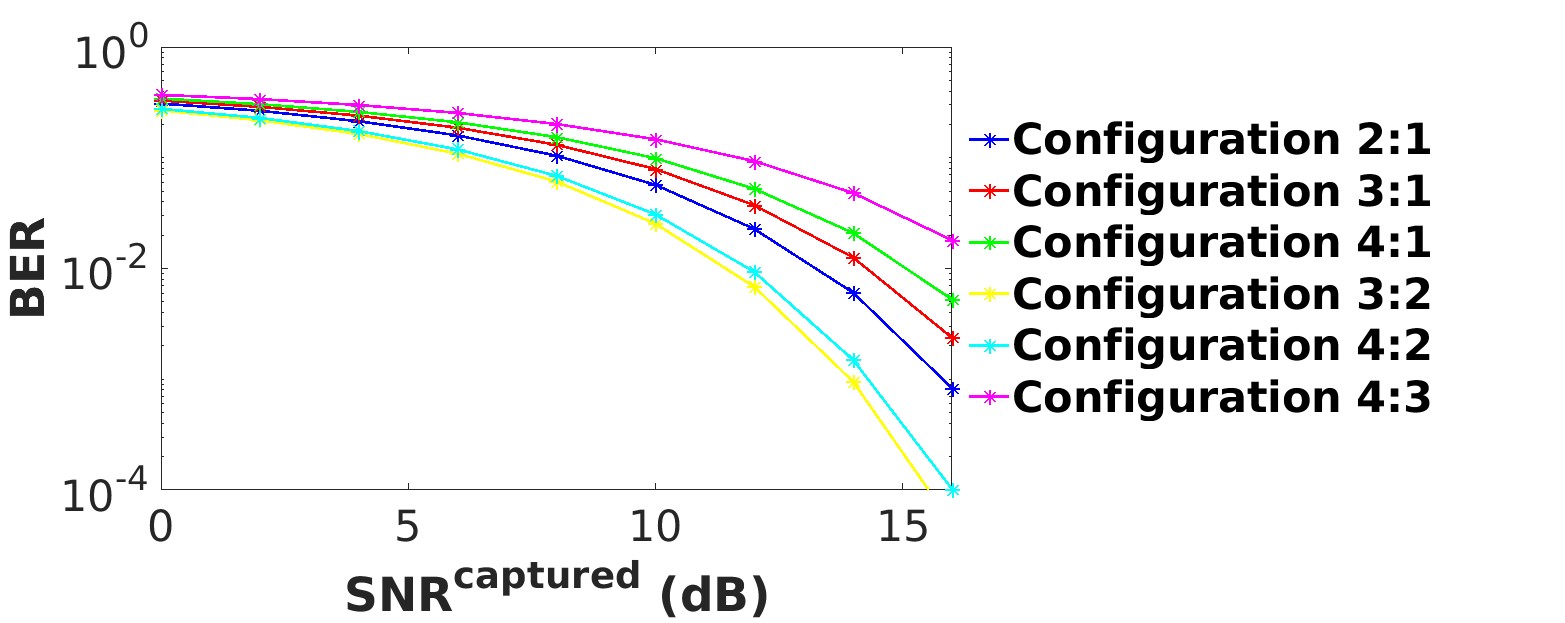}}
\caption{Fig. 19.	BER rate for a the different configurations of the PCS  of the XPOL antenna.\label{fig20}}
\end{figure}

\begin{figure}
\centerline{\includegraphics[width=3.5in]{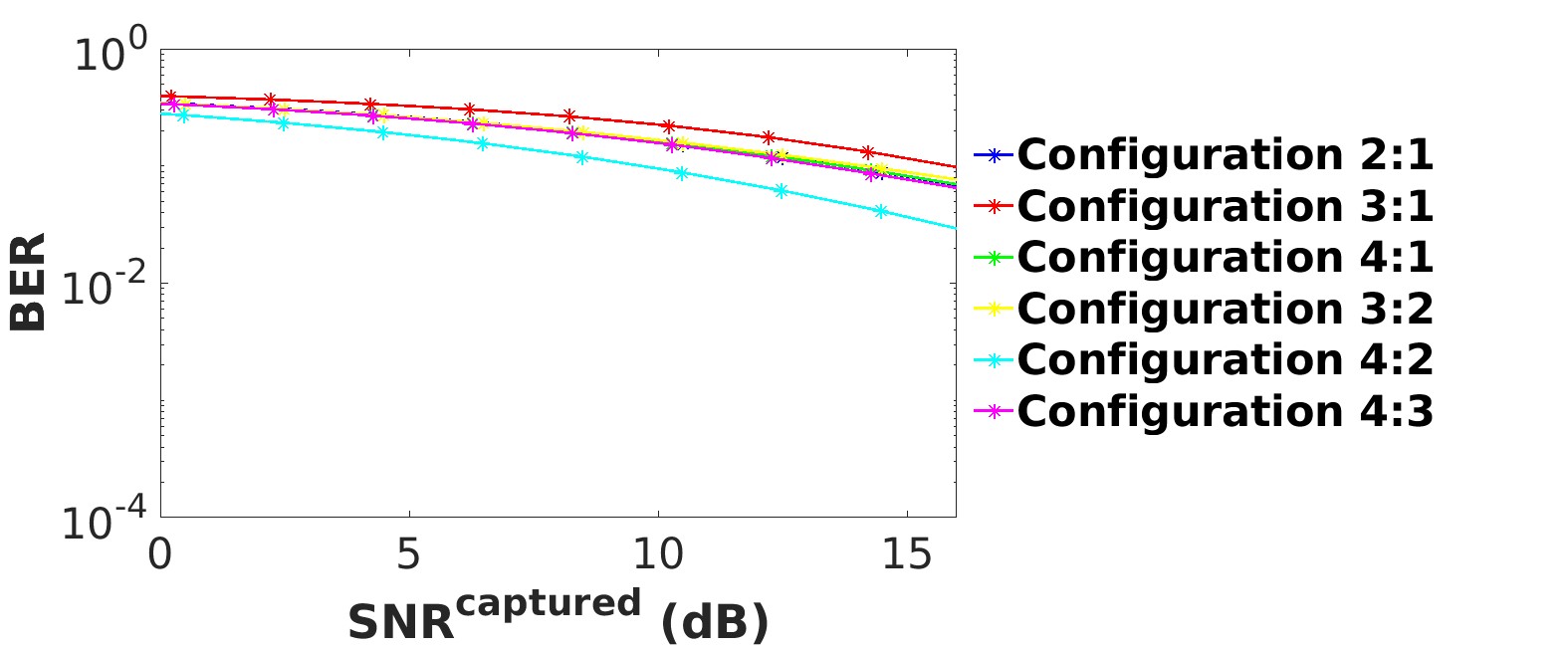}}
\caption{Fig. 20.	BER rate for a the different configurations of the PCS of the 4PR antenna.\label{fig21}}
\end{figure}

In this section, we evaluate numerically the BER as a function of the $SNR^{captured}$. The results are reported in Figs. \ref{fig19}, \ref{fig20}, \ref{fig21}.
From the figures, we evince that the performance depends on the selected polarization patterns. For example, it is easier for the LSE detector to detect the bits if the two patterns are uncorrelated. 

From Fig. \ref{fig19}, we observe that the best configuration for the SRR corresponds to the states 2 and 1, which have an envelope correlation coefficient lower than 0.1 \cite{ref9}. Also, the configuration with the states 3 and 2 is one of the worst and has a correlation coefficient close to 0.5 \cite{ref9}. 

From Fig. \ref{fig20}, we observe that the BER performance of the XPOL depends on the configuration of the chosen states. The best configuration is obtained for the states 3 and 2, and for the states 4 and 2. In both cases, we observe that these states have opposite pattern directions (see Fig. \ref{fig15}). This property, in fact, reduces the correlation between the patterns and the states are more easily detectable. 

From Fig. \ref{fig21}, we observe that the performance of the 4PR tag is worse compared to the XPOL and the SRR tags. This can be explained by the fact that the patterns of the dipole antennas are more correlated. We can note that the configuration with the states 4 and 2 has the best performance because these two states correspond to the best orientations of the dipole (see Section \ref{sectionIIIA}) and because these states are cross-polarized (Fig. \ref{fig18}).

In conclusion, our analysis shows that the SRR antenna is the best antenna, thanks to its low inter-patterns correlation properties.

\section{Conclusion}
\label{sectionVII}
In this paper, we have shown the advantages of using a polarization reconfigurable tag in ambient backscatter systems. Such a tag takes advantage of different polarizations in order to improve the robustness of the communication against the direct source-to-reader interference. With the aid of analysis, simulations, and experiments, we have shown that a realistic  
4 polarization reconfigurable tag provides performance that is similar to an ideal tag with 80 polarization states.

\section*{Acknowledgment}
The authors would like to thank K. Rachedi for his support.

\begin{IEEEbiographynophoto}{R. FARA} received the master's degree of electrical engineering in 2018 from the National Institute of Applied Sciences of Lyon, France. He is currently pursuing the Ph.D. degree with Orange Labs Networks, Chatillon, France in collaboration with the laboratory of signals and systems, CentraleSupelec, University of Paris-Saclay, Paris, France. His current research interest includes ambient backscatter communications.
\end{IEEEbiographynophoto}
\begin{IEEEbiographynophoto}{D.-T. PHAN-HUY} received the degree in engineering from Supelec, in 2001, and the Ph.D. degree in electronics and telecommunications from the National Institute of Applied Sciences of Rennes, France, in 2015. In 2001, she joined France Telecom R\&D (now Orange Labs Networks), Chatillon, France. She led the national French collaborative research projects TRIMARAN (2011-2014) and SpatialModulation (2016-2019). She participated to the following 5G PPP projects: METIS, Fantastic 5G, mmMAGIC and 5GCAR. She holds more than 40 patents and has published more than 40 papers. She is the recipient of several awards in France: “Prix Impact Economique des Rencontres du Numérique 2016” from the French National Research Agency, “Grand Prix de l’Electronique du Général Ferrié 2018” from the French Society of Electricity, Electronics and Information and Communication Technologies and the “Prix Irène Joliot Curie 2018 – categorie Femme-Recherche-Entreprise” from the French Ministry of Education and Research. Her research interests include wireless communications and beamforming, spatial modulation, predictor antenna, backscattering and intelligent reflecting surfaces.
\end{IEEEbiographynophoto}
\begin{IEEEbiographynophoto}{A. OURIR} received his engineering degree from ENIT (Tunis Tunisia) in 2003. He received his Ph.D. Degree in Physics from Paris Sud University (Orsay France) in 2007. Since 2008, he is a CNRS research engineer at Institut Langevin (ESPCI), Paris, France. He has developed original devices based on passive and active metamaterials for antennas. His current research interests include metamaterial based antennas, electromagnetic subwavelength imaging and wave propagation in artificial materials. He is involved in SpatialModulation project.
\end{IEEEbiographynophoto}
\begin{IEEEbiographynophoto}{Y. KOKAR} received the Master’s degree in Engineering from the Montpellier university graduate engineering school, Montpellier, France, in 2004. In 2006, he joined the Electronics and Telecommunications Institute of Rennes (IETR) Laboratory at INSA of Rennes, as a Research Engineer. His research interests lie at the intersection of communication theory, wireless networks and their implementation through hardware and software defined.
\end{IEEEbiographynophoto}
\begin{IEEEbiographynophoto}{J.-C. PREVOTET} obtained his Phd in 2003 from the UPMC. He is currently an associate professor at IETR/INSA de Rennes. His major interests are embedded and reconfigurable systems and real time systems in general. In particular, his applicative subjects deal with communication systems and the way to optimize their architecture onto real platform. He is also deeply involved in the real time management of these communication platforms under the supervision of an embedded operating system
\end{IEEEbiographynophoto}
\begin{IEEEbiographynophoto}{M. HELARD} received her engineering and PhD degrees from INSA of Rennes, France, in 1981 and 1984, respectively and her habilitation degree in 2004. After being with Orange Labs as research engineer, she joined INSA of Rennes, France as a full professor in 2007 where she led the IETR department. She was involved in several collaborative research projects. Her research interests include wired and wireless communications and MIMO techniques.
\end{IEEEbiographynophoto}
\begin{IEEEbiographynophoto}{M. DI RENZO} was born in L’Aquila, Italy, in 1978. He received the Laurea (cum laude) and Ph.D. degrees in electrical engineering from the University of L’Aquila, Italy, in 2003 and 2007, respectively, and the Habilitation a Diriger des Recherches (Doctor of Science) degree from University Paris-Sud, France, in 2013. Since 2010, he has been with the French National Center for Scientiﬁc Research (CNRS), where he is a CNRS Research Director (CNRS Professor) in the Laboratory of Signals and Systems (L2S) of Paris-Saclay University – CNRS and CentraleSupelec, Paris, France. He is a Nokia Foundation Visiting Professor at Aalto University, Helsinki, Finland, and a Honorary Professor at University Technology Sydney, Sydney, Australia. He serves as the Editor-in-Chief of IEEE Communications Letters. He is a Distinguished Lecturer of the IEEE Vehicular Technology Society and IEEE Communications Society. He is a Highly Cited Researcher according to Clarivate Analytics and Web of Science, a Fellow of the IEEE, and a recipient of several research awards.
\end{IEEEbiographynophoto}
\begin{IEEEbiographynophoto}{J. DE ROSNY} received the M.S. Degree and the Ph.D. degree from the University UPMC, Paris, France in 1996 and 2000, respectively, in wave physics. He was a postdoctoral at Scripps Research Institute, California, USA, in 2000-2001. In 2001, he joined CNRS at Laboratoire Ondes et Acoustique, France. Since 2014, he is a CNRS senior scientist at Institut Langevin, Paris, France. His research interests include telecommunications in complex media, acoustic and electromagnetic waves based imaging.
\end{IEEEbiographynophoto}
\end{document}